\documentclass[pop,fleqn]{w-art}
\usepackage{times}
\usepackage{w-thm}
\usepackage{amssymb}
\usepackage{cite}
\usepackage[]{graphicx}
\begin{document}
\DOIsuffix{theDOIsuffix}
\Volume{51}
\Issue{1}
\Month{01}
\Year{2003}
\pagespan{1}{}



\title[D-brane model building]{Progress in D-brane model building}


\author[F.~Marchesano]{Fernando G. Marchesano%
  \footnote{Corresponding author\quad E-mail:~\textsf{marchesa@theorie.physik.uni-muenchen.de}}}
\address[]{Arnold Sommerfeld Center for Theoretical Physics\\
Ludwig-Maximilians Universit\"at\\
Theresienstra\ss e 37\\
80333 M\"unchen}
\begin{abstract}
The state of the art in D-brane model building is briefly reviewed, focusing on recent achievements in the construction of $D=4$ ${\cal N} = 1$ type II string vacua with semi-realistic gauge sectors. Such progress relies on a better understanding of the spectrum of BPS D-branes, the effective field theory obtained from them and the explicit construction of vacua. We first consider D-branes in standard Calabi-Yau compactifications, and then the more involved case of  compactifications with fluxes. We discuss how the non-trivial interplay between D-branes and fluxes modifies the previous model-building rules, as well as provides new possibilities to connect string theory to particle physics.
\end{abstract}
\maketitle                   





\section{Motivation}

As often emphasized in the literature, string theory gained its popularity among physicists in the mid-80's, as it was recognized as an outstanding candidate for a unified theory of all forces in nature. Not only would it incorporate quantum gravity, but also supersymmetry and non-Abelian gauge interactions. With such ingredients it was soon realized that, upon compactification, one could conceive $D=4$ ${\cal N} = 1$ effective theories remarkably close to natural extensions of the Standard Model \cite{GSW2}.

Further developments of the theory showed that the set of string theory vacua is quite complex even at a qualitative level. On the one hand we learned that seemingly different vacua could be dual to each other and, in fact, should be considered to be the same. On the other hand, it has been estimated that the set of semi-realistic vacua is enormous. This latter point has recently raised the question of whether string theory can reproduce almost any $D=4$ effective theory that one may think of and, in particular, many possible physics beyond the Standard Model. If that was the case, one may drop the classical strategy to focus on a particular string theory model by means of some vacuum selection principle, and instead try to extract physical information out of the ensemble of realistic vacua via an statistical approach \cite{Douglas03}.

Now, while string theory has proven to be an excellent generator of phenomenologically interesting scenarios, a string vacuum reproducing the physics observed at particle accelerators is yet to be found. Individual vacua analyzed so far not only fail to describe fully realistic $D=4$ physics because of some technical details like extra massless particles or unrealistic couplings, but also because of deep conceptual issues like gauge coupling unification, hierarchy problems and supersymmetry breaking. Moreover, recent cosmological observations have slightly changed our perspective on how a semi-realistic vacuum should be obtained and, in general, agreement with cosmological data puts even more constraints on realistic string constructions. 

Pointing out these facts is not intended to discourage those interested in the field of model building.  On the contrary, there has recently been huge progress (some of which the present review is based on) in constructing vacua which are more and more realistic, and we expect much more progress to come. It is however important to keep in mind that some work is still needed before string theory can claim to reproduce everyday physics as an effective theory. Whether such realistic description arises from a single vacuum or a myriad of them we still do not know, but in any case it would be highly desirable to have explicit examples of realistic vacua before applying anthropic, statistical or vacuum selection principles to string theory. In addition, pushing our knowledge of string theory in order to obtain constructions which are more and more realistic may unveil new unexpected characteristics that may affect our present understanding of the theory and, in particular, our way of connecting string theory to particle physics. After all, it is the quest for moduli stabilization via flux compactifications what shaped the statistical approach to string theory as we know it today.

The purpose of this note is to describe some recent developments in the field of model building, and in particular in those string constructions aiming to reproduce the Standard Model as an effective theory. Just like in the original picture described in \cite{chsw85}, we will focus on weakly coupled string theory vacua yielding ${\cal N} = 1$, $D=4$ Poincar\'e invariant effective field theories. Unlike in \cite{chsw85}, however, we will not consider the heterotic scenario, but rather type II orientifold compactifications. Type II theories are not more fundamental than any other string theory and neither should they be any better than, say, heterotic strings in order to reproduce $D=4$ semi-realistic physics. However, it is in the framework of type II orientifold models where most developments regarding moduli stabilization and vacua statistics has been performed and so, if one would like to understand model building results in the context of the vacuum degeneracy problem, this is the place to look.

Notice that if we are interested in reproducing the Standard Model from type II string theories closed strings are not enough \cite{dkv87}, and hence we should consider D-branes \cite{D-branes}. D-branes are the best understood non-perturbative objects in string theory and, in fact, they played a central role when the string web of dualities was unveiled. Since then, more and more interesting properties of these dynamical objects have been discovered, which have in turn given rise to phenomenologically interesting scenarios like, e.g., brane worlds \cite{braneworlds} and brane inflation \cite{braneinflation}. In addition, since D-branes localize gauge interactions and matter fields charged under them, one may build type II vacua by steps in a `bottom-up' approach, as opposed to the `top-down' strategy implicit in \cite{chsw85}. As shown in \cite{aiqu00}, such bottom-up/modularity philosophy allows for more flexibility when building semi-realistic vacua, and it is particularly effective for models of D3-branes at singularities. The same philosophy can be applied to most semi-realistic D-brane constructions and, in fact, it is present in all of the models discussed below. Finally, D-branes can be naturally combined with some other model building ideas arising in type II theories, like those based on AdS/CFT \cite{Maldacena98}, warped throats \cite{ks00} and compactifications with background fluxes, to which we will dedicate the last part of this review.

For all of the above reasons and probably further ones, the field of D-brane model building has quickly evolved and it is by now quite a mature subject. While the purpose of this note is to briefly report on some recent developments, there are more extensive as well as pedagogical reviews in the D-brane literature \cite{reviews} and also on the subject of flux compactifications \cite{reviewsflux}, to which we refer the reader for further details. Finally, although here we will focus on D-brane models applied to reproduce the Standard Model of particle physics, the subject of D-brane cosmology has significantly developed in the past few years. For overviews on this rich subject, we refer the reader to \cite{reviewscosmo}.

\section{Type IIA model building}\label{typeIIAmb}

While historically the last ones to be explored, chiral type IIA vacua have become extremely popular in the past few years, and are in fact the main source of chiral D-brane models in the recent literature. One of the reasons for their popularity comes from the dictionary between geometry and effective field theory that exists in any D-brane model, and which is particularly simple for intersecting D6-branes.

\subsection{Intersecting brane worlds}

Let us first describe how $D=4$, ${\cal N} = 1$ chiral models can be obtained in the type IIA framework. As we aim for a four-dimensional effective theory of (quantum) gravity, we first compactify type IIA string theory on a six-dimensional manifold ${\bf X}_6$, which for the time being we assume smooth and large enough so that the supergravity approximation is valid. In order to include a $U(N)$ gauge theory in this setup we can simply consider $N$ type IIA $D(2p)$-branes, filling up the four non-compact dimensions and wrapping a $2p - 3$ cycle $\Pi_{2p-3} \subset {\bf X}_6$. Since we are assuming ${\bf X}_6$ to be compact so will be $\Pi_{2p-3}$, and we will obtain a $D=4$, $U(N)$ gauge theory upon standard Kaluza-Klein reduction.

Imposing $D=4$ supersymmetry will constrain the topology and geometry of this general setup. Let us initially set every RR and NSNS background flux to zero, condition which will be relaxed later on in section \ref{fluxes}. Then, supersymmetry in the closed string sector imposes the four non-compact dimensions to be Poincar\'e invariant, and the six internal dimensions to form a Ricci-flat, K\"ahler manifold. Otherwise speaking, in order to achieve a supersymmetric closed string sector we consider the ten-dimensional metric ansatz $M_4 \times {\bf X}_6$, where $M_4$ stands for the four-dimensional Minkowski background and ${\bf X}_6$ is a Calabi-Yau three-fold. One can then check that massless content of the theory will be given by a $D=4$, ${\cal N} = 2$ supergravity multiplet, the dilaton hypermultiplet, $h^{2,1}$ further hypermultiplets and $h^{1,1}$ vector multiplets.

In general, this extended ${\cal N} = 2$ supersymmetry will be broken when introducing space-filling D-branes on top of the above metric background.\footnote{When introducing space-filling D-branes, the initial ansatz of a Calabi-Yau manifold without background fluxes is strictly speaking no longer true, since a D-brane will be a source for the metric, dilaton and RR field strengths. However, in the weak coupling limit which we are considering, such D-brane backreaction can in principle be neglected, and hence for model building purposes we can work in the Calabi-Yau, fluxless limit. Of course, once that a complete, consistent model is constructed, one should be able to take into account the D-brane backreaction in order to recover the full supergravity background.} Since at this level of the construction we still want to preserve some supersymmetry, we should consider D-branes which are $\frac{1}{2}$-BPS states of the full, ten-dimensional theory. This will directly imply that the previous $U(N)$ gauge symmetry will be promoted to a full $U(N)$ super Yang-Mills theory.  

Which is the full set of space-filling BPS D-branes in a type IIA compactification? As we will see in section \ref{coisotropic}, the answer is more involved than one may initially think. However, we can already go quite far in type IIA model building from naive considerations. Indeed, recall that type IIA space-filling $D(2p)$-branes need to wrap odd-dimensional cycles $\Pi_{2p-3}$ in the internal space ${\bf X}_6$ and that, in a generic Calabi-Yau three-fold, the Betti numbers are $b_1 = b_5 = 0$. This means that the only chance to wrap a D-brane on a non-trivial $(2p-3)$-cycle (in de Rham homology) is to set $p=3$. We should then consider D6-branes wrapping 3-cycles $\Pi_3 \subset {\bf X}_6$, whose supergravity BPS charge will be classified by the homology class $[\Pi_3]$. For simplicity, let us consider a single D6-brane, carrying a $U(1)$ gauge group. The BPS conditions for such a D-brane are \cite{mmms99}
\begin{eqnarray}
\label{FIIA}
{\cal F} + iJ &  = & 0 \quad \quad {\rm F-flatness}\\
\label{DIIA}
{\rm Im}\, e^{-i\theta}\Omega & = & 0 \quad \quad {\rm D-flatness}
\end{eqnarray}
where ${\cal F} = 2\pi \alpha' F + B$ is the usual gauge invariant field strength in a D-brane worldvolume,   while $J$ and $\Omega$ are bulk forms that can be obtained from the internal Killing spinor of the compactification. In particular, for a Calabi-Yau three-fold, $J$ is the K\"ahler 2-form and $\Omega$ is the holomorphic 3-form, and both are closed and non-degenerate. In all of the expressions of this kind the bulk $p$-forms like $B$, $J$ and $\Omega$ are implicitly pull-backed to the internal D-brane worldvolume, in this case the 3-cycle $\Pi_3$. Finally, $e^{-i\theta}$ is a constant phase which only depends on the homology class of $\Pi_3$.

One can now interpret geometrically such supersymmetry conditions. On the one hand, the F-flatness condition (\ref{FIIA}) implies that a BPS D6-brane can only host a flat $U(1)$ bundle on its internal worldvolume (i.e., it only admits pure Wilson lines) and that the pull-back of $J$ must vanish. In the language of symplectic geometry, this latter condition is rephrased as $\Pi_3$ being a Lagrangian submanifold of ${\bf X}_6$. The D-flatness condition (\ref{DIIA}) will further restrict the D6-brane embedding, implying that $\Pi_3$ must be a special Lagrangian submanifold (sLag), and so its volume can be computed by means of the integral
\begin{equation}
{\rm Vol}(\Pi_3)\, =\, \int_{\Pi_3} e^{-i\theta}\Omega
\label{volume}
\end{equation}
which by (\ref{DIIA}) is a real number, and we can always choose $e^{-i\theta}$ such that it is also positive. Manifolds whose volume can be computed by integrating certain non-degenerate, closed $p$-forms are dubbed calibrated submanifolds, and were first introduced in \cite{hl82} in order to describe volume-minimizing objects in special holonomy manifolds. When translated into physics, all this implies that a D6-brane in a fixed homology class $[\Pi_3]$ (i.e., fixed BPS charges) minimizes its volume (i.e., its energy) by wrapping a special Lagrangian submanifold. This is indeed the kind of property that we would expect from a BPS-like object in a supersymmetric theory \cite{Gauntlett03}.

\begin{figure}[htb]
\includegraphics[width=\linewidth, height=2cm]{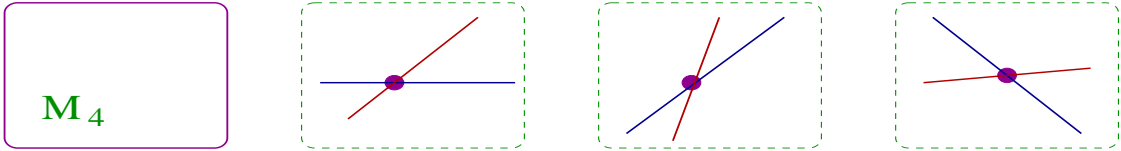}
\caption{Two flat D6-branes intersecting at angles.}
\label{atangles}
\end{figure}

How do we get chirality in this framework? The easiest way to see this is to consider two flat D6-branes, $D6_a$ and $D6_b$, embedded in flat ten-dimensional space. As these two D6-branes fill-up four common spacetime dimensions, we can write their worldvolume as $D6_{a,b} \simeq M_4 \times \Pi_3^{a,b}$, where $\Pi_3^{a,b}$ are each a 3-plane in ${\mathbb{R}}^6$. Generically, these two 3-planes will intersect transversally in a point $p \in {\mathbb{R}}^6$ (see figure \ref{atangles}) and, by the analysis of \cite{bdl96}, we know that the spectrum of open strings stretching between $D6_a$ and $D6_b$ (or $D6_aD6_b$ sector) will contain a massless chiral fermion localized at $M_4 \times \{p\}$. The matrix rotating $\Pi_3^a$ to $\Pi_3^b$ will give us further information about this spectrum in the $D6_aD6_b$ sector. The orientation of the rotation matrix, either positive or negative, will give us the chirality of the $D=4$ fermion. The eigenvalues, which can be understood as three rotation angles, will give us the spectrum of light scalars. In particular, for a rotation that can be embedded in $SU(3)$ we will obtain a massless scalar, which together with the chiral fermion will complete a $D=4$, ${\cal N} = 1$ chiral multiplet. This signals the fact that two D6-branes related by an $SU(3)$ rotation will preserve a common ${\cal N} = 1$ supersymmetry.

Notice that a 3-plane is the simplest example of sLag in a rather trivial (and non-compact) Calabi-Yau manifold. In general we will have to consider a pair of D6-branes wrapping complicated 3-cycles $\Pi_3^a$ and $\Pi_3^b$ on a smooth, curved and compact Calabi-Yau manifold ${\bf X}_6$. However, if such submanifolds intersect transversally, we can approximate the geometry near the  intersection by figure \ref{atangles}, and hence the previous discussion still applies. Since our D6-branes now wrap compact submanifolds, they give rise to a $D=4$ gauge group $U(N_a) \times U(N_b)$, where $N_{a,b}$ are the number of D6-branes wrapped on each 3-cycle. The chiral fermion at such intersection then transforms in the bifundamental representation $(N_a, \overline{N}_b)$ and, because  $\Pi_3^a$ and $\Pi_3^b$ are no longer 3-planes, they can intersect several times. Notice that each intersection will contribute with a chiral fermion in the same bifundamental representation, but that the chiralities can be different for each intersection. One should then consider the net chiral spectrum in the $D6_aD6_b$ sector, given by the signed intersection number
\begin{equation}
I_{ab}\, =\, [\Pi_3^a] \cdot [\Pi_3^b]
\label{intersection}
\end{equation}
which is in fact a topological invariant, in the sense that it only depends on the homology classes of the 3-cycles wrapped by the D6-branes.

\begin{figure}[tb]
\includegraphics[width=12cm, height=6.5cm]{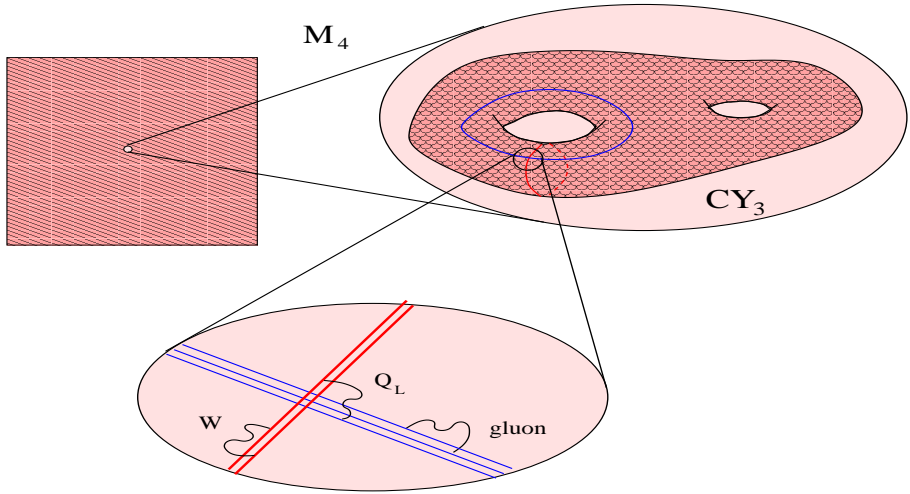}
\caption{Intersecting brane worlds.}
\label{interworld}
\end{figure}

All these phenomenologically interesting features led, in \cite{bgkl00,afiru00}, to propose the scenario where a realistic $D=4$ effective field theory is obtained from intersecting D6-branes, as summarized in figure \ref{interworld}. In such an scenario $D=4$ gravity is obtained from the closed string sector of type IIA string theory when compactified on a six-dimensional manifold ${\bf X}_6$, $D=4$ $U(N)$ gauge theories are obtained from a stack of $N$ D6-branes wrapping 3-cycles $\Pi_3 \subset {\bf X}_6$ and, finally, chiral fermions in bifundamentals are obtained from the intersection of two different stacks of D6-branes. Because two 3-cycles can intersect several times, we have a natural, geometrical mechanism to reproduce the replication of chiral families observed in nature.

Notice that this scenario is a specific realization of the brane world idea, where gravity and gauge fields propagate in different dimensions of the internal space-time. This opens up the possibility of considering realistic constructions based on large extra dimensions and supersymmetry explicitly broken at a low string scale \cite{afiru00,cim02c,Kokorelis02,Uranga02}. Nevertheless, we will restrict our scenario to models with ${\cal N} = 1$ SUSY at the string scale, which requires ${\bf X}_6$ to be a Calabi-Yau manifold, each 3-cycle $\Pi_3$ to be a sLag and, if we want each D6-brane to preserve the same ${\cal N}=1$ supersymmetry inside the ${\cal N}=2$ bulk superalgebra, the same calibration phase $e^{i\theta}$ in (\ref{DIIA}) for each $D6$-brane in the model \cite{Douglas01}.

While the above scenario looks quite promising, one may wonder how much effort does one need to build a semi-realistic model and obtain all the physical information out of it. After all, almost no Calabi-Yau metrics are known explicitly, and also very few examples of special Lagrangians have been constructed in the mathematical literature. However, one of the advantages of the above setup is that, given relatively few topological data, one can already extract the chiral spectrum of the theory. Indeed, if we specify {\it i)} the number $N_a$ of D6-branes wrapping a 3-cycle $\Pi_3^a$, {\it ii)} the homology class $[\Pi_3^a] \subset H_3({\bf X_6}, \mathbb{R})$ of each 3-cycle and {\it iii)} the intersection numbers (\ref{intersection}) between each pair of 3-cycles, the $D=4$ chiral spectrum is given by table \ref{specIIA}.
\begin{table}
\caption{Chiral spectrum of an intersecting D6-brane model. $U(1)_a$ stands for the trace generator of $U(N_a)$.}
\label{specIIA}
\begin{tabular}{@{}lll@{}}
\hline
Non-Abelian gauge group & $\prod_a SU(N_a)$\\
Massless $U(1)$'s &  $\sum_a c_a U(1)_a$ such that $\sum_a c_a [\Pi_3^a] = 0$ \\
Chiral fermions & $\sum_{a<b}\, I_{ab} (N_a, \overline{N}_b)$\\
\hline
\end{tabular}
\end{table}
This implies that,  in order to know if an MSSM-like spectrum can be obtained from a specific Calabi-Yau, we do not need to know its metric or the explicit embedding of the 3-cycles $\Pi_3^a$, but just the homology classes of these 3-cycles in ${\bf X_6}$ and the intersection product between them. Of course, in order to claim that we have constructed a model we should also prove that in each homology class $[\Pi_3^a]$ that we are considering there exists a representative which is a special Lagrangian 3-cycle, and this is in general quite a complicated problem \cite{Thomas01}. Nevertheless, the above topological information already tells us which are the 3-cycles that one should look at, and this is already a great deal of simplification.

\subsection{Effective field theory}

Still at this abstract level of the discussion, one can extract plenty of information regarding the low energy effective action. First, the gauge kinetic function of a D6-brane wrapping a sLag $\Pi_3^a$ is given by \cite{cim02a,bbkl02}
\begin{equation}
f_a\, =\, \frac{1}{(4\pi^2\alpha')^{3/2}}  \int_{\Pi_3^a}e^{-\phi}\, {\rm Re} \left(e^{-i\theta}\Omega\right) + i C_3
\label{gkfIIA}
\end{equation}
where $e^{-\phi}$ stands for the $D=10$ dilaton, and $C_3$ is the RR 3-form potential. From this expression one can see that the gauge kinetic function will only depend on the complex structure moduli of ${\bf X}_6$, as we would expect from the general results on Calabi-Yau compactifications \cite{bdlr99}. Of course, this expression for the gauge kinetic function is only valid at tree-level, and it is expected to be corrected at one-loop. Such threshold corrections are quite hard to compute in general, as one needs to know the full massive string spectrum of the compactification. As a result, they have so far only been computed in simple cases like when ${\bf X}_6$ is a toroidal orbifold \cite{threshold}. 

In general the open strings in the $D6_aD6_a$ sector do not give rise to a pure $U(N_a)$ super Yang-Mills theory, but also to a set of massless, chiral multiplets $\Phi_a^r$ in the adjoint representation. These adjoint fields parameterize the local moduli space of a BPS D6-brane, which is given by a smooth manifold of complex dimension $b_1(\Pi_3^a)$, i.e., the number of harmonic 1-forms in $\Pi_3^a$ \cite{McLean98}. Such moduli space is a naive geometrical approximation, and it may be partially lifted by a stringy superpotential generated by open string world-sheet instantons \cite{instantons}. 

\begin{figure}[htb]
\includegraphics[width=70mm,height=40mm]{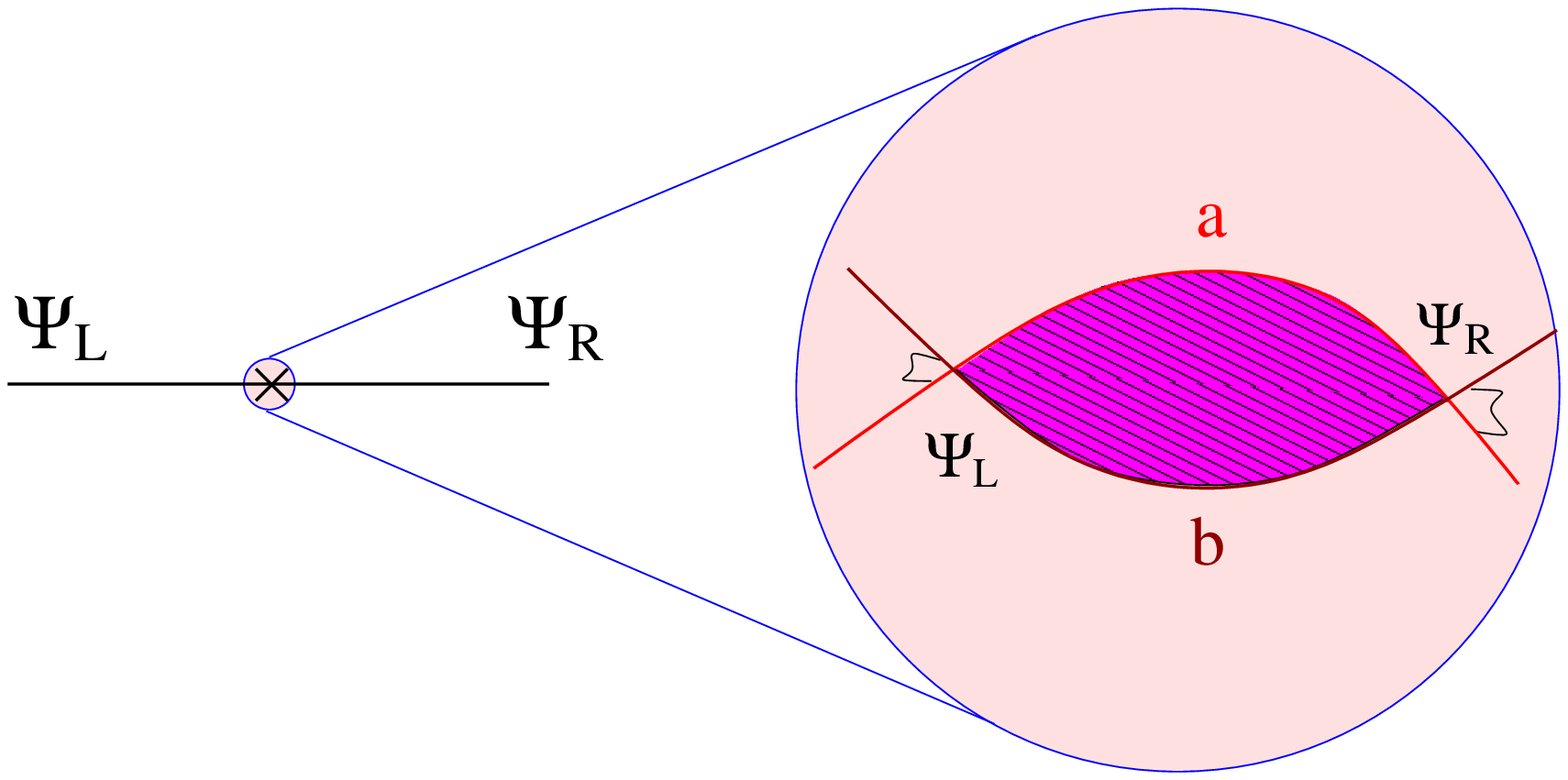}~a)
\hfil
\includegraphics[width=70mm,height=40mm]{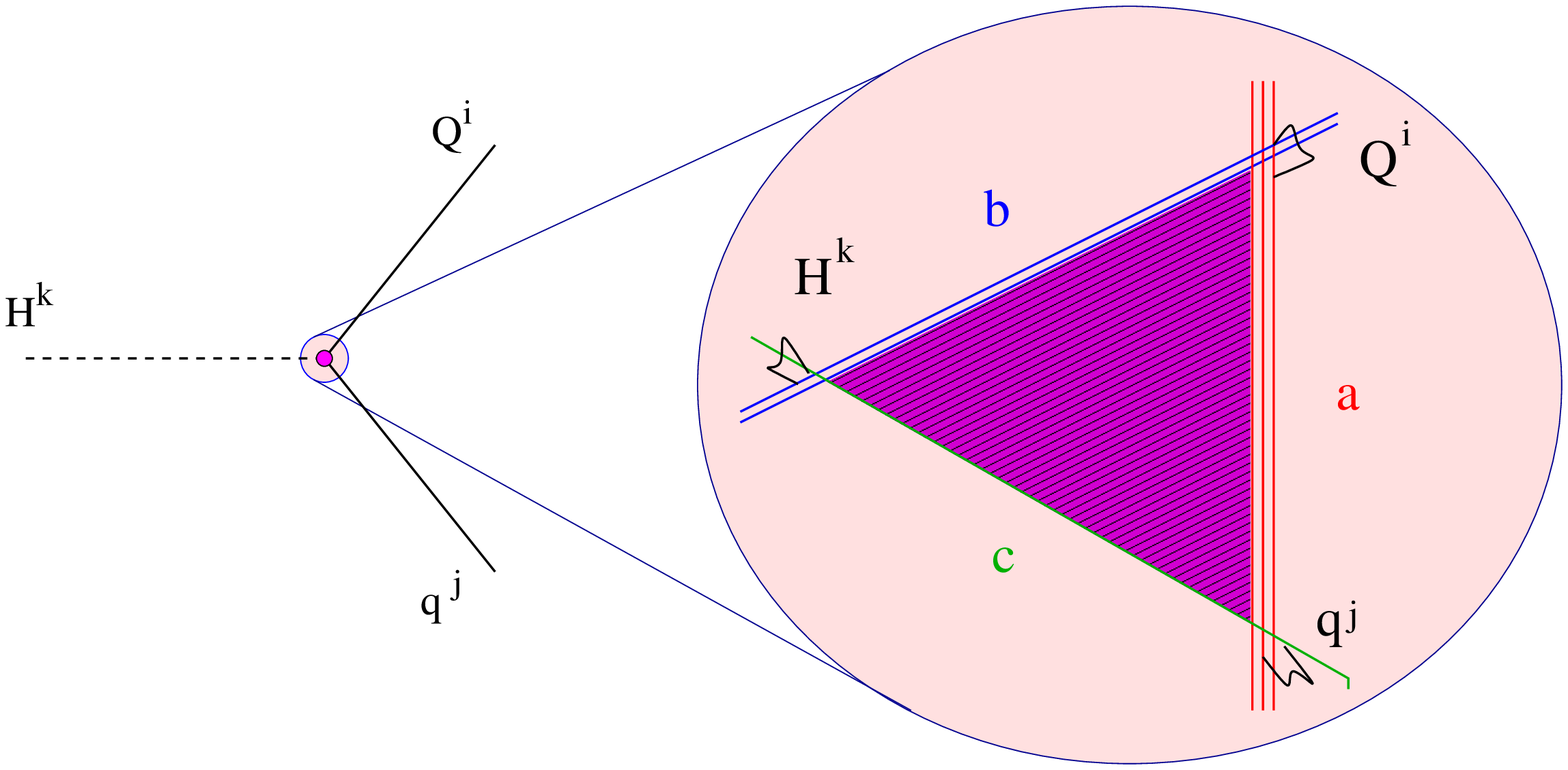}~b)
\caption{Worldsheet instantons as generators of {\em a)} masses for vector-like pairs and {\em b)} Yukawa couplings.}
\label{ws}
\end{figure}

World-sheet instantons contributing to a D6-brane superpotential will be holomorphic or antiholomorphic surfaces with the topology of a disk. When dealing with a single D6-brane, the instanton boundary lies on a non-trivial 1-cycle inside $\Pi_3^a$, and it contributes to the above mentioned superpotential for the adjoint multiplets. However, when considering a full model with several D6-branes, further world-sheet instanton generated superpotentials will arise. Indeed, let us consider the case where two D6-branes, $D6_a$ and $D6_b$, intersect twice, first with positive orientation and then with negative orientation. This means that a world-sheet instanton can connect both intersections, its boundary lying half in $D6_a$ and half in $D6_b$ (see figure \ref{ws}.a). From the point of view of the effective theory, the $D6_aD6_b$ sector will contain two chiral fermions of opposite chirality, with a mass term generated by the world-sheet instanton connecting them. Finally, when considering systems of three intersecting D6-branes, world-sheet instantons will generate Yukawa couplings between chiral fields (see figure \ref{ws}.b), as originally noticed in \cite{afiru00}. 

While conceptually quite simple, the actual computation of world-sheet generated superpotentials can be quite complicated in generic Calabi-Yau compactifications, and it is a rich mathematical subject of current research \cite{Fukaya}. One can nevertheless perform such computation in the case where the compactification manifold ${\bf X}_6$ is a toroidal orbifold and the D6-branes wrap flat submanifolds $\Pi_3 \simeq T^3$. In that case the superpotential for adjoint fields is trivial and two D6-branes always intersect with the same orientation, so the main superpotential terms to be computed are the Yukawa couplings between chiral fields. Such computation has been performed in \cite{cim03}, and the results can be summarized as follows. Let us consider that our compactification manifold is ${\bf X}_6 = (T^2)_1 \times (T^2)_2 \times (T^2)_3$ and that each D$6_\alpha$-brane, $\alpha = a, b, c$, is wrapping a 3-cycle of the form $\Pi_3^\alpha = ({\rm 1-cycle})_1 \times ({\rm 1-cycle})_2 \times ({\rm 1-cycle})_3$. Assuming this basic setup, on which most of the intersecting D-brane models are based, we obtain a Yukawa coupling of the form
\begin{eqnarray}
\label{Yukphys}
Y_{ijk} & = & e^{K/2} (K_{ab}K_{bc}K_{ca})^{-1/2}\, W_{ijk}\\
\label{YukIIA}
W_{ijk} & = & \sigma_{abc} 
\prod_{r=1}^3 \vartheta
\left[
\begin{array}{c}
\delta_{ijk}^{r} \\ 0
\end{array}
\right]\,
(\Phi^r, I_r T_r)
\end{eqnarray}
where $r$ labels each $(T^2)_r$ of the compactification, and $i, j, k$ label the intersection points of the $D6_aD6_b$, $D6_bD6_c$ and $D6_cD6_a$ sectors. Here $\vartheta$ is the standard Jacobi theta function with characteristics, $\Phi^r$ is a linear combination of the vev's of the open string moduli, $T_r = B_r + iA_r$ is the complexified K\"ahler modulus of the $r^{\rm th}$ 2-torus, $I_r$ is a product of intersection numbers, and $\sigma_{abc} = {\rm sign} (I_{ab} I_{bc} I_{ca})$. Finally, $\delta_{ijk}^{r}$ encodes the (rather simple) only flavor dependence of the Yukawa coupling. To this formula one should add the prescription that, for some choices $i, j, k$ of chiral fields, there is no world-sheet instanton and hence the corresponding Yukawa coupling vanishes. We refer the reader to \cite{cim03,cim04} fur further details on these results.

Notice that $W_{ijk}$ will contain the full flavor dependence of the Yukawa coupling and it is then the source of potentially interesting textures in, e.g., quark mass matrices. However, the physical Yukawa coupling to be measured also depends on the K\"ahler potential via the standard supergravity formula (\ref{Yukphys}). Unlike the superpotential, the K\"ahler potential will not be a holomorphic quantity, and it will depend on both K\"ahler and complex structure moduli of the compactification manifold. This makes it a much more difficult quantity to compute, specially when charged chiral matter is involved. Particularly interesting is the K\"ahler metric for chiral matter $K_{\alpha\beta}$, again computed for intersecting D6-branes in flat space \cite{lmrs04}:
\begin{equation}
\kappa^2 K_{\alpha\beta}\, \propto\, e^{\phi_4} \prod_{r=1}^3 ({\rm Im\, } T_r)^{-\varphi_{\alpha\beta}^r} \sqrt{\frac{2\pi\, \Gamma(\varphi_{\alpha\beta}^r) }{\Gamma(1-\varphi_{\alpha\beta}^r)}}
\label{KahlerIIA}
\end{equation}
where $\phi_4$ is the four-dimensional dilaton and $\varphi_{\alpha\beta}^r$ are the intersection angles on $(T^2)_r$, normalized such that $0 < \varphi_{\alpha\beta}^r < 1$, $\sum_r \varphi_{\alpha\beta}^r = 1 = \varphi_{ab}^r + \varphi_{bc}^r + \varphi_{ca}^r$. This is however just a partial result, since it assumes a specific point on the D6-brane moduli space, namely ${\rm Im\, }\Phi^r = 0$ and, by the results of \cite{cim03,cim04}, we would expect a non-trivial dependence of the K\"ahler metric $K_{\alpha\beta}$ on ${\rm Im\, }\Phi^r$. In any case, already a great deal of physical information can be obtained from (\ref{KahlerIIA}) which is crucial in, e.g., the soft term computations of section \ref{fluxes}. Finally, in the same framework of flat D6-branes in flat space, one can perform exact CFT computations of four-point \cite{cp03,ao03a} and higher N-point functions \cite{ao03b} of chiral fields at the intersections, which are relevant in order to discuss phenomenological issues like FCNC \cite{ams03} or proton decay \cite{kw03,cr06}.

Another effective field theory quantity that can be discussed in generality are Fayet-Iliopoulos terms, which will generate a D-term potential. Indeed, recall that if two D6-branes $D6_a$ and $D6_b$, are wrapped over special Lagrangians with different calibration phases then they will preserve different ${\cal N} = 1$ subalgebras of ${\cal N} = 2$, and so the combined system will break supersymmetry. Since the calibration phase only appears in the D-term condition (\ref{DIIA}), we would expect such supersymmetry breaking to appear as an FI-term in our effective theory. Indeed,  as was observed in \cite{km99}, depending on the relative calibration phase $\theta_b - \theta_a$ the pair of D6-branes will recombine into a single one, will preserve supersymmetry or will remain as it stands. By computing the spectrum of light scalars at the intersection one can see that these situations correspond to the lightest scalar being be tachyonic, massless or massive \cite{afiru00}. Such FI-terms were explicitly computed in \cite{cim02a} (see also \cite{bbkl02,vz06}), obtaining
\begin{equation}
\xi_{ab} = \frac{\theta_a - \theta_b}{2\alpha'}
\label{FItermIIA}
\end{equation}
for the FI-term felt by an open string in the $D6_aD6_b$ sector, hence charged under $U(1)_{ab} = U(1)_b - U(1)_a$. These terms indeed reproduce the spectrum of open string scalars mentioned above, as well as the scalar potential computed in \cite{bklo01}.

\subsection{Orientifolding}

Our discussion on model building can so far be considered as `local', in the sense that it has only involved one, two, or at most three D-branes, but not the full set of D-branes needed to construct a complete model. When building such a model in a full-fledged compactification, one should worry about a set of consistency conditions known as RR tadpole cancellation conditions. From the supergravity perspective, RR tadpole conditions are understood from the fact that D-branes are sources of RR fields \cite{Polchinski95} and so, in order for these fields to be well-defined in a compact space ${\bf X}_6$, the total D-brane charge needs to vanish. For intersecting D6-brane models, such consistency conditions can be simply expressed as $\sum_a N_a\, [\Pi_3^a] = 0$, where $N_a$ and $ [\Pi_3^a]$ are defined as above \cite{afiru00}.

It turns out that, in Calabi-Yau compactifications to $D=4$ Minkowski vacua, tadpole cancellation conditions, D-branes and supersymmetry are incompatible features. One can see this by noting that RR conditions imply that $\sum_a N_a \int_{[\Pi_3^a]} e^{-i\theta_a} \Omega = 0$, where we have used the fact that any Calabi-Yau is a complex manifold and hence $d\Omega = 0$. Supersymmetry implies that each phase $\theta_a$ is equal and that each term of this sum is a positive real number, namely the tension of each individual D-brane. Hence, we are requiring a sum of D-brane tensions to vanish, which can only be satisfied by having no D-branes at all.

The way out is to add new objects into the theory, which preserve supersymmetry but carry negative tension. This can be achieved by means of a variation of the above setup, known as orientifold quotient, by which different states in our worldsheet theory are identified to be the same. More precisely, we mod out our type IIA configuration by $\Omega \sigma$, where $\Omega$ is the usual worldsheet orientation reversal and $\sigma$ is a $\mathbb{Z}_2$ symmetry of the compactification manifold ${\bf X}_6$. Such orientifold quotient will modify the closed string sector of our previous theory and, for instance, it will break the previous ${\cal N}=2$ bulk supersymmetry. In order to preserve an ${\cal N}=1$ supersymmetry in the closed string sector, one should require $\sigma$ to be an antiholomorphic involution of ${\bf X}_6$. The massless closed string spectrum then arranges into $D=4$, ${\cal N}=1$ multiplets as one gravity multiplet, $h_+^{1,1}$ vector multiplets and $h_-^{1,1} + h^{2,1} +1$ chiral multiplets (here $h_\pm^{1,1}$ refer to the eigenspaces of $H^{1,1}({\bf X}_6)$ induced by the action of $\sigma$), as can be verified by explicit dimensional reduction \cite{gl04a}.

Orientifold quotients are important for model building because, from the point of view of $D=10$ supergravity, they create non-dynamical negative tension objects, known as orientifold planes \cite{Sagnotti87}.  An O$p$-plane is a source of the metric, dilaton and RR field strengths just like a D$p$-brane, but it contributes with the opposite sign to the Bianchi identities and the equation of motion.\footnote{This may look quite disturbing from a geometrical perspective, since near the O$p$-plane the metric becomes negative definite. However, one should have in mind that our supergravity intuition is just an approximation of the full string theory, valid only in certain cases. For instance, an O6-plane can be lifted to M-theory and described geometrically as an Atiyah-Hitchin manifold \cite{O6}.} In the particular setup above, the orientifolded theory incorporates one or several space-filling O6-planes wrapping 3-cycles $\Pi_{O6},$ defined by the fixed point set of $\sigma$. Since they source the RR field strength with (four times) the opposite charge as a D6-brane would do, the RR tadpoles now become
\begin{equation}
\sum_a N_a [\Pi_3^a]\, =\, 4\, [\Pi_{O6}]
\label{RRIIA}
\end{equation}
which implies that space-filling D6-branes are now required by the theory in order to have a consistent compactification. Moreover, because $\sigma$ is antiholomorphic, $\Pi_{O6}$ will be a special Lagrangian submanifold of fixed phase which, without loss of generality, we will take to be $+1$. This means that, after orientifolding, the calibration phase of a BPS D6-brane in (\ref{DIIA}) is fixed by the closed string background to be $\theta = 0$, which can be seen as a consequence of the bulk theory being ${\cal N}=1$ instead of ${\cal N}=2$.

Orientifolding a string theory will not only change the closed string sector but also the open string massless spectrum. First, since $\Omega\sigma$ must be a symmetry of the D-brane sector, for each D6$_a$-brane wrapping a 3-cycle $\Pi_3^a$, we must include the orientifold image D6$_{a'}$, wrapping $\Pi_3^{a'} = \sigma(\Pi_3^a)$. If $\Pi_3^{a'} = \Pi_3^a$ then the gauge group becomes $USp(2N_a)$ or $SO(2N_a)$. If $\Pi_3^{a'} \neq \Pi_3^a$, then the gauge group is still $U(N_a)$, but the spectrum of open strings is now more complicated, as summarized in table \ref{specIIAori} \cite{bgkl00,imr01,csu01}.
\begin{table}[htb]
\caption{Chiral spectrum of an intersecting D6-brane orientifolded model, where for simplicity we are assuming that there are no D6-branes with $SO(2N_a)$ or $USp(2N_a)$ gauge group. $H_3^+({\bf X}_6, \mathbb{R})$ stands for the space of 3-cycle homology classes of ${\bf X}_6$ fixed by $\sigma$. 
}
\label{specIIAori}
\begin{tabular}{@{}lll@{}}
\hline
Non-Abelian gauge group & $\prod_a SU(N_a)$\\
Massless $U(1)$'s &  $\sum_a c_a U(1)_a$ such that $\sum_a c_a [\Pi_3^a] \in H_3^+({\bf X}_6, \mathbb{R})$ \\
Chiral fermions & $\sum_{a<b}\, I_{ab} (N_a, \overline{N}_b) \, + \, I_{ab'} (N_a, N_b) $\\
& $\frac{1}{2} (I_{aa'} - I_{a,O}) {\bf Sym}_a\, + \, \frac{1}{2} (I_{aa'} + I_{a,O}) {\bf Antisym}_a$\\
\hline
\end{tabular}
\end{table}

\subsection{Explicit models}

The above discussion shows that, by considering intersecting D6-branes, one has all the basic ingredients to conceive semi-realistic string vacua. However, in order to measure how good the Intersecting Brane World proposal is, one needs to build explicit models where all these phenomenologically interesting features are put to work.

Let us give a simple example of MSSM-like model based on intersecting D6-branes, following \cite{cim03,cim02d,ms04}. First, we choose our D6-brane content to be given by table \ref{guay}, where two of the D6-branes are wrapping 3-cycles $\Pi_3^b$, $\Pi_3^c$ fixed by the orientifold involution $\sigma$, and such that their gauge group is $USp(2) \simeq SU(2)$. We also choose $\Pi_3^a$ and $\Pi_3^d$ to lie in the same homology class, up to a torsion element. Second, let us impose the intersection numbers $I_{ab} = - I_{ac} = 3$ and $I_{aa'} = 0$. Then, using the information of table \ref{specIIAori}, we see that the gauge group of this model is given by $SU(3) \times SU(2)_L \times SU(2)_R \times U(1)_{B-L}$, that is, a left-right extension of the Standard Model gauge group. If $\Pi_3^a$ and $\Pi_3^d$ not only lie in the same homology class but are in fact the same 3-cycle, then the gauge group will be further extended to the $SU(4) \times SU(2)_L \times SU(2)_R$ Pati-Salam model \cite{ps74}. In addition, the choice of intersection numbers yields precisely 3 families of quarks and leptons with the appropriate quantum numbers, and no extra chiral exotics.

\begin{table}[htb]
\caption{Pre-model of intersecting D6-branes. $\Pi_3^a$ and $\Pi_3^{d}$ may in general be different but we require $[\Pi_3^a] = [\Pi_3^{d}]$.}
\label{guay}
\begin{tabular}{@{}llcccc@{}}
\hline
D-brane content & & $3 \Pi_3^a$ & $\Pi_3^b$ & $\Pi_3^c$ & $\Pi_3^{d}$\\
Gauge group & & $SU(3) \times U(1)_a$ & $SU(2)$ & $SU(2)$ & $U(1)_d$ \\
\hline
\end{tabular}
\end{table}

Of course, this choice of topological data should only be seen as a `pre-model' of intersecting D6-branes. In order to claim that we have built such an MSSM-like model, we need to construct an explicit realization in terms of a Calabi-Yau orientifold and four sLags with the required intersection numbers. Once such a realization has been achieved, one should check that all the consistency conditions like RR tadpole cancellation are satisfied and, if not, extra BPS D6-branes should be added in order to satisfy them. This part of the problem is technically more difficult, but it is quite important in order to understand the phenomenological possibilities of our general approach. Indeed, finding an explicit model not only serves as a proof of concept that the pre-model of table \ref{guay} (or any other) can be realized in string theory, but also gives important information about quantities beyond the chiral spectrum of the theory: the Higgs sector, the precise value of the gauge kinetic functions and Yukawa couplings, etc. Since these less robust quantities are also the least understood in the Standard Model, one would like to see if anything can be learnt from the string theory perspective. Finally, only by explicit constructions can one address even more subtle questions like moduli stabilization, supersymmetry breaking and hierarchy problems in the present context.

\begin{figure}[htb]
\includegraphics[width=10cm, height=3.5cm]{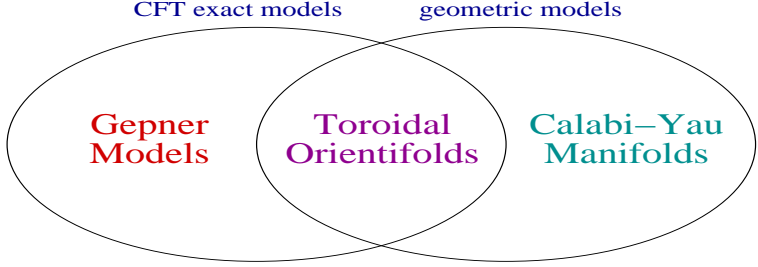}
\caption{Models discussed in the text in terms of geometrical and 2D CFT properties.}
\label{venn}
\end{figure}

So let us provide an explicit realization of the above pre-model, based on a toroidal orientifold. Toroidal orientifolds are a particularly simple class of Calabi-Yau orientifolds, where the metric background is of the form $T^6/\Gamma$ and $\Gamma$ is a discrete subgroup of $SU(3)$. The advantage of these models is that not only do they follow the geometric intuition of Calabi-Yau constructions, but they also allow for CFT exact computations and they are free of $\alpha'$ corrections. As illustrated in figure \ref{venn}, such geometric, CFT-exact models provide the best of both worlds in terms of the D-brane vacua explored up to now, so they are the obvious place to start testing our scenario.

Many of these toroidal orientifolds can be built by first considering a factorized geometry of the form $(T^2)_1 \times (T^2)_2 \times (T^2)_3$, which must in addition be compatible with the antiholomorphic involution $\sigma\, :\, z^i \mapsto \bar{z}^i$, ($z^i$ is the complex coordinate on the $i^{\rm th}$ $T^2$). This background has plenty of well-known special Lagrangian submanifolds, which are of the form $({\rm 1-cycle})_1 \times ({\rm 1-cycle})_2 \times ({\rm 1-cycle})_3$, $({\rm 1-cycle})_i \subset (T^2)_i$, so they can be easily described in terms of integer numbers as
\begin{equation}
\Pi_3^a\, =\, (n_a^1, m_a^1) \times (n_a^2, m_a^2) \times (n_a^3, m_a^3)
\label{3cycleT6}
\end{equation}
whereas the intersection number between two of them is given by
\begin{equation}
I_{ab}\, =\, [\Pi_3^a]\cdot[\Pi_3^b]\, =\, (n_a^1m_b^1-m_a^1n_b^1) \times (n_a^2m_b^2-m_a^2n_b^2) \times (n_a^3m_b^3-m_a^3n_b^3)
\label{interT6}
\end{equation}
Notice that (\ref{3cycleT6}) does not contain the full set of special Lagrangians of $(T^2)_1 \times (T^2)_2 \times (T^2)_3$, but it is already quite a big set and it has the nice property that the associated D6-branes can be treated as exact boundary states of the CFT. In addition, in order to construct sLags of calibration phase zero, we just need to impose the condition $\theta_1 + \theta_2 + \theta_3 = 0\, {\rm mod}\, 2\pi$, where $\theta_i$ is the angle between $(n_a^1, m_a^1)$ and ${\rm Re}\, z^i$.

Finally, we are interested in backgrounds of the form $T^6/\Gamma$, so one should implement the orbifold projection in both closed and open string sectors.\footnote{If $\Gamma$ is trivial one cannot achieve supersymmetric models \cite{bgkl00}.} This will in general change the spectrum of (fractional) D6-branes and the computation of their intersection numbers although, in the case of $\Gamma = \mathbb{Z}_2 \times \mathbb{Z}_2$ which we will consider here, one can see that the expressions (\ref{3cycleT6}) and (\ref{interT6}) do not change. We refer the reader to \cite{csu01} for an explanation of this point as well as other model building subtleties associated to this background. 

\begin{table}[htb]
\caption{Explicit ${\cal N} = 1$ D6-brane model on $T^2 \times T^2 \times T^2/\mathbb{Z}_2 \times \mathbb{Z}_2$ realizing the pre-model of table \ref{guay}.}
\label{Ymodel}
\begin{tabular}{@{}|llccc|@{}}
\hline
$N_\alpha$  & & $(n_\alpha^{1},m_\alpha^{1})$  &  $(n_\alpha^{2},m_\alpha^{2})$
&  $(n_\alpha^{3},m_\alpha^{3})$ \\
\hline $N_a = 3 + 1$ & & $(1,0)$ & $(3,1)$ & $(3,-1)$  \\
$N_b=1$ & & $(0,1)$ &  $ (1,0)$  & $(0,-1)$ \\
$N_c=1$ & & $(0,1)$ &  $(0,-1)$  & $(1,0)$  \\
\hline
$N_{h_1}= 1$ & & $(-2,1)$  & $(-3,1)$ & $(-4,1)$ \\
$N_{h_2}= 1$ & & $(-2,1)$ & $(-4,1)$ & $(-3,1)$ \\
$N_f = 20$ & & $(1,0)$ &  $(1,0)$  & $(1,0)$  \\
\hline
\end{tabular}
\end{table}

The explicit model which realizes the pre-model of table \ref{guay} is presented in table \ref{Ymodel}. Here, the upper box of the table corresponds to the set of D6-branes of table \ref{guay} (we have set $\Pi_3^a = \Pi_3^d$ for simplicity), and the lower box are extra D6-branes that are needed for the model to be consistent.\footnote{We refer the reader to \cite{csu01} for the explicit RR tadpole cancellation conditions of this background. In addition to such consistency conditions, there are further ones invisible to a supergravity analysis, and which are of pure K-theoretical nature \cite{Uranga00}. For the orientifold background at hand, such conditions were computed in \cite{ms04}.} If we focus on the upper set of D6-branes, seen as an MSSM-like sector of the theory, we can compute the Higgs sector of this model, which arises from $D6_bD6_c$ open strings. We then find just the MSSM Higgs content, together with a non-vanishing $\mu$-term. The computation of the Yukawa couplings was performed in \cite{cim03}, with the intriguing result that, while no coupling is in principle forbidden, only one family of quarks and leptons is massive. Finally, the gauge kinetic functions of this model satisfy some interesting ratios, which have been analyzed in the context of gauge coupling unification in \cite{bls03}. 

If we now also consider the lower set of D6-branes we see that they contribute non-trivially to the massless spectrum. First, because of their presence the gauge group is enlarged to
\begin{equation}
SU(4) \times SU(2)_L \times SU(2)_R \times U(1)' \times USp(40)
\label{gaugeYmodel}
\end{equation}
where $U(1)'= U(1)_a/4 - 2 (U(1)_{h_1} - U(1)_{h_2})$. Second, they will create chiral matter charged both under the Pati-Salam gauge group and this extra $U(1)$ factor. This is indeed quite undesirable phenomenologically, but nothing that cannot be solved by looking at a more general set of models, as we will show below. 

In any case, this simple example is still far from being fully realistic. For instance, the D6-brane wrapping $\Pi_3^a$ has three complex moduli which, from the point of view of the effective theory, are seen as three adjoint multiplets spoiling $SU(3)$ confinement. In addition, while having a $\mu$-term is quite interesting phenomenologically, there is no preferred value for such $\mu$-term, so we are faced with the usual $\mu$-problem of MSSM-like scenarios. Since the $\mu$-term of this model is nothing but the vev of some D6-brane moduli, it turns out that both problems are related to the moduli problem, now in the D-brane sector of the theory. Hence one should try to solve them in the context of moduli stabilization, which we will explore in section \ref{fluxes} via the addition of background fluxes. As we will comment, such flux scenario could also give some interesting input for more fundamental issues like supersymmetry breaking.

While relatively simple, this is of course not the only ${\cal N} = 1$ semi-realistic vacuum based on intersecting D6-branes, and there is a quite extensive literature on the subject. The first examples of type IIA vacua with D6-branes at angles were also based on toroidal orientifolds \cite{bgk99,fhs00}, although these were automatically non-chiral models. It was then realized in \cite{csu01} that one could build ${\cal N} = 1$, chiral vacua by placing the D6-branes not on top of the orientifold planes, and still arranging things in order to satisfy all the consistency conditions. Such chiral constructions were based on the orbifold group $\Gamma = \mathbb{Z}_2 \times \mathbb{Z}_2$, the one considered above, although further research showed that one could build ${\cal N} = 1$ chiral models for more involved orbifold backgrounds \cite{Honecker03,bgo02,ho04,bcms05,bp06}.

On the other hand, given the general discussion presented above one expects that the intersecting brane worlds proposal can be implemented in general Calabi-Yau backgrounds, and not only in toroidal orientifolds. Some non-trivial steps in this direction were given in \cite{Uranga02}, which addressed the construction of non-compact Calabi-Yau geometries involving intersecting D6-branes. These non-compact models should be seen as a small region of a bigger, compact Calabi-Yau which is the full closed string background. The local patch described by the non-compact geometry contains the 3-cycles where the D6-branes realizing the Standard Model would be located, whereas some other regions could give rise to hidden sectors of the theory. These local constructions are quite natural in a bottom-up model building approach, and we will encounter them again in the next section, when analyzing models of D3-branes at singularities.

In principle, one can apply the same set of ideas to even more general orientifold constructions, like those that are well-defined CFT constructions but do not have a clear geometric interpretation. Indeed, another set of ${\cal N} = 1$ chiral models are those based on Gepner orientifolds, whose model building rules were developed in \cite{gepner}. Semi-realistic models based on these backgrounds were constructed in \cite{aaj04,dhs04,adks06}, using the same strategy described here to construct the model in table \ref{Ymodel}. In fact, the computational possibilities of these Rational CFT's allowed, in \cite{dhs04,adks06}, to perform quite an extensive search of explicit Gepner constructions based on semi-realistic pre-models. It was obtained the intriguing result that the pre-model of table \ref{guay} (or close relatives) yields the most frequent class of semi-realistic constructions. These constructions also showed that, unlike the case of the model in table \ref{Ymodel}, in many cases the minimal D-brane content of table \ref{guay} is enough to construct a fully consistent model, and that no extra D-branes creating chiral exotics are needed. We would expect this and some other Gepner-based statements to apply to general Calabi-Yau constructions, which do however remain to be explored.

\section{Type IIB model building}

Naively, type IIB chiral vacua look richer than those of type IIA, simply because there are more classes of D-branes that can be used to build a model. Indeed, we can now consider D3, D5, D7 and D9-branes as space-filling objects, wrapping even-dimensional cycles ${\cal S}_{2n}$ in a compact manifold ${\bf Y}_6$. In addition, for ${\bf Y}_6$ a Calabi-Yau background the D-brane BPS conditions \cite{mmms99,dfr00,bhw05b}
\begin{eqnarray}
\label{FIIB}
{\cal S}_{2n}\, {\rm is\, holomorphic},\, \quad {\cal F}^{(2,0)} =  0 & \quad &{\rm F-flatness}\\
\label{DIIB}
{\rm Im}\quad e^{-i\theta} e^{{\cal F} + i J} {\sqrt{\hat{A}({T_{{\cal S}_{2n}}})}}  \, =\,  0 & \quad & {\rm D-flatness}
\end{eqnarray}
allow for a non-trivial, holomorphic gauge bundle ${\cal F}$. Here $\hat{A}$ is the A-roof genus encoding the D-brane curvature couplings and $T_{{\cal S}_{2n}}$ is the tangent bundle of ${\cal S}_{2n}$. The phase $e^{i\theta}$ has the same interpretation as in (\ref{DIIB}) and it will also be fixed by an orientifold projection. However, the two values that $\theta$ can take look now very different geometrically. The phase $\theta = 0$ corresponds to type IIB compactifications with $O3$ and $O7$-planes, whereas $\theta = \pi/2$ applies to backgrounds with $O5$ and/or $O9$-planes, including type I vacua. The closed string spectrum of these compactifications is summarized in table \ref{closedtypeIIB} (see \cite{gl04b} for further details).\footnote{The orientifold quotient is again of the form $\Omega\sigma$, but now $\sigma$ is a holomorphic involution fixing holomorphic submanifolds. Its action on $\Omega$ and $J$ is given by $\sigma(\Omega) = \Omega$ for $\theta = \pi/2$, $\sigma(\Omega) = - \Omega$ for $\theta = 0$ and $\sigma(J) = J$ in both cases.}
\begin{table}[htb]
\caption{Closed string spectrum of type IIB compactified on a Calabi-Yau, before and after the orientifold projections. The numbers $h^{p,q}_\pm$ correspond to the usual Hodge numbers, split with respect to the action of the orientifold quotient.}
\label{closedtypeIIB}
\begin{tabular}{@{}lllcl@{}}
\hline
Theory & SUSY & & $D=4$ spectrum & \\
\hline
Type IIB & ${\cal N} = 2$ & gravity multiplet,& $h^{2,1}$ vector mult.,& $h^{1,1} + 1$ hypermult. \\
Type IIB with O3/O7's& ${\cal N} = 1$ & gravity multiplet,& $h^{2,1}_+$ vector mult.,& $h^{1,1} + h^{2,1}_- + 1$ chiral mult. \\
Type IIB with O5/O9's& ${\cal N} = 1$ & gravity multiplet, & $h^{2,1}_-$ vector mult.,& $h^{1,1} + h^{2,1}_+ + 1$ chiral mult. \\
\hline
\end{tabular}
\end{table}

Besides more kinds of D-branes and orientifold projections, there seem to be more mechanisms in order to produce $D=4$ chirality in type IIB models. The simplest one is to consider two D9-branes with different gauge bundles, ${\cal F}_a$ and ${\cal F}_b$, so that open strings in the $D9_aD9_b$ sector feel the difference ${\cal F}_{ab} = {\cal F}_{a} - {\cal F}_{b} = 2\pi \alpha' (F_a - F_b)$ of `open string fluxes'. From a field theory perspective, this will modify the internal Dirac operator and produce $D=4$ chiral fermions via Kaluza-Klein reduction of $D=10$ gauginos \cite{Bachas95}, in an analogous way to the heterotic models of \cite{chsw85}. Another mechanism to create chirality is to place D3-branes at orbifold singularities \cite{dm96,dgm97}, so that the ${\cal N}=4$ spectrum of $N$ D3-branes in smooth space is projected out to an ${\cal N}=1$ chiral spectrum. Finally, it is easy to see that any pair of D-branes that intersects point-like in the internal dimensions yields a $D=4$ chiral fermion from each intersection. This not only applies to the pair of D6-branes discussed before, but also to a D3 and a D9-brane (which intersect at the D3-brane position) or to a D5 and a D7-brane intersecting transversely. 

This higher diversity of type IIB chiral models is somehow suspicious since, by mirror symmetry, we expect the $D=4$ physics of any type IIB vacuum to be also reproduced by some type IIA model. Rather than something fundamental, the differences between type IIB models should come from our supergravity perspective, which makes a D3-brane and a D9-brane to appear as very different objects. In principle, both kinds of type IIB models can be mapped to type IIA compactifications with intersecting D6-branes, such that the dictionary of table \ref{mirror} is satisfied. Hence, one should be able to give a unified description of type IIB model building, just like it was done for type IIA in the previous section.
\begin{table}[htb]
\caption{Mirror map between type IIA compactified on ${\bf X}_6$ and type IIB on ${\bf Y}_6$, in terms of their Hodge numbers.}
\label{mirror}
\begin{tabular}{@{}ll@{}}
\hline
Type IIA $\leftrightarrow$ type IIB & $h^{1,1}({\bf X}_6) = h^{2,1}({\bf Y}_6)$,\, $h^{2,1}({\bf X}_6) = h^{1,1}({\bf Y}_6)$ \\
Type IIA with O6's $\leftrightarrow$ type IIB with O3/O7's& $h^{1,1}_\pm ({\bf X}_6) = h^{2,1}_\pm ({\bf Y}_6)$,\, $h^{2,1}_\pm ({\bf X}_6) = h^{1,1}_\pm ({\bf Y}_6)$ \\
Type IIA with O6's $\leftrightarrow$ type IIB with O5/O9's& $h^{1,1}_\pm ({\bf X}_6) = h^{2,1}_\mp ({\bf Y}_6)$,\, $h^{2,1}_\pm ({\bf X}_6) = h^{1,1}_\mp ({\bf Y}_6)$ \\
\hline
\end{tabular}
\end{table}

An important hint for such unified picture is that the different ways of producing chirality in type IIB are not unrelated. First, the chiral spectrum of D3-branes at singularities is understood in terms of fractional D3-branes, which are the minimal objects that are well-defined in this singular geometry. However, after blowing up the singularity, we realize that fractional D-branes are higher dimensional D-branes wrapping a collapsed 2 or 4-cycle, and with some particular gauge bundle ${\cal F}$ on them. Second, a D9-brane with a non-trivial gauge bundle ${\cal F}$ (usually dubbed magnetized D9-brane) can be seen as a D9-brane where lower dimensional D7, D5 and D3-branes have been dissolved \cite{Douglas95}. The computation of the chiral spectrum between two magnetized D9-branes can then be understood as the sum of intersection numbers between their elementary components.\footnote{Since intersection numbers are defined in (co)homology, they are well-defined even if a D-brane is dissolved or smoothed-out.}

In general, one can describe a type IIB D-brane as a bound state of several D3, D5, D7 and D9-branes, where the 2 and 4-cycles wrapped by the D5 and D7-branes should also be specified. This information is naturally encoded in the sum of even-dimensional forms \cite{ghm96,cy97,mm97}
\begin{equation}
Q^a\, =\, ch(F_a) \, \sqrt{\hat{A}(T_{{\cal S}_{2n}^a})/\hat{A}(N_{{\cal S}_{2n}^a})} \, \delta ({\cal S}_{2n}^a)
\label{chargeIIB}
\end{equation}
where $\delta ({\cal S}_{2n}^a)$ is a $(6-2n)$-form with $\delta$-function support in ${\cal S}_{2n}^a$, such that $[{\cal S}_{2n}^a]$ and $[\delta ({\cal S}_{2n}^a)]$ are  Poincar\'e dual to each other in ${\bf Y}_6$,\footnote{Physically, $\delta ({\cal S}_{2n})$ is the source term for the RR field-strength: $dF_{5-2n} = \delta ({\cal S}_{2n})$.} $ch(F_a)$ is the Chern character of the gauge bundle $F_a$ on the D-brane, and $N_{{\cal S}_{2n}^a}$ is the normal bundle of ${\cal S}_{2n}^a$. The charge of such D-brane is given by the cohomology class $[Q^a] = H^0 \oplus H^2 \oplus H^4 \oplus H^6$, where $H^i$ stands for $H^i({\bf Y}_6, \mathbb{R})$, and where we should also impose charge quantization. That is, type IIB D-brane charges are vectors in a $2 (1 + h^{1,1}({\bf Y}_6))$-dimensional lattice, which nicely fits with type IIA charges taking values in a lattice of $b_3({\bf X}_6) = 2(1 + h^{2,1}({\bf X}_6))$ dimensions and the mirror map of table \ref{mirror}. 

The parallelism between type IIA and type IIB models can be further pushed, in the sense that we can also compute the chiral spectrum between two type IIB D9-branes from an intersection number between two vector charges, as
\begin{equation}
I_{ab}\, =\, [Q^a] \cdot [Q^b]\, = \, \int_{{\bf Y}_6} Q^a \wedge (Q^b)^* \, =\,  \int_{{\bf Y}_6} ch(F_a) \, ch(-F_b) \, \hat{A}({\bf Y}_6)
\label{intersectionIIB}
\end{equation}
where $(Q^b)^* = ch(-F_b) \, \sqrt{\hat{A}({\bf Y}_6)}$. When dealing with D-branes of lower dimensions, one should basically compute the integral of $ch(F_a) \, ch(-F_b)$ on the intersection manifold ${\cal S}_{2n}^a \cap {\cal S}_{2n'}^b$. One can still express the chiral spectrum in terms of an intersection product (\ref{intersectionIIB}), but the definition of the conjugate vector $(Q^b)^*$ is then more subtle. In such cases things are better formulated by describing D-branes as coherent sheaves, and obtaining the intersection numbers by computing $Ext$ groups among them \cite{sheaves}.

The bottom-line is that, with only a few topological data given by (\ref{chargeIIB}), one can compute the chiral spectrum of a type IIB Calabi-Yau D-brane model by using the simple rules of table \ref{specIIA} and replacing $[\Pi_3^a]$ by $[Q^a]$. The same applies to the RR consistency conditions, which now read $\sum_a N_a [Q^a] = 0$. Finally, one can easily generalize table \ref{specIIAori} and eq.(\ref{RRIIA}) to the two classes of type IIB orientifolds, and so the model building rules and concepts for type IIA and type IIB orientifold vacua are basically the same.

Finally, let us stress that this discussion, based on the supergravity approximation to string theory, misses some important characteristic of D-brane RR charges, which is the fact that they are classified by K-theory, rather than by (co)homology \cite{mm97,Witten98}. While this extends the spectrum of D-branes in a given compactification, the difference only involves torsional RR charges, which do not contribute to intersection products like (\ref{intersection}) or (\ref{intersectionIIB}). Hence, this subtle point does not affect the computation of the chiral spectrum of our compactification. On the other hand, it is important in order to build a consistent model, since in general it gives rise to extra consistency constraints beyond (\ref{RRIIA}) \cite{Uranga00}.

Just like for type IIA, one can compute the most relevant $D=4$ quantities in the effective action of a type IIB model. In general, we expect the same kind of results at both sides of the mirror map. This is easy to see for some quantities like the gauge kinetic function, which in the type IIB case reads \cite{bhw05b,hklpz06}
\begin{equation}
f_a = \int_{{\cal S}_{2n}^a}\hspace{-.25cm} \left( e^{-\phi}\, {\rm Re}\, \left(e^{-i\theta} e^{iJ}\right) + i \sum_k C_{2k}\right) \, e^{\cal F} \, \sqrt{\frac{\hat{A}(T_{{\cal S}_{2n}^a})}{\hat{A}(N_{{\cal S}_{2n}^a})}}  = \int_{[\tilde{Q}^a]} \hspace{-.25cm} e^{-\phi}\, {\rm Re} \left(e^{-i\theta} e^{B + iJ} \right) + i \sum_k C_{2k}\, e^{B}
\label{gkfIIB}
\end{equation}
where $[\tilde{Q}^a]$ stands for the Poincar\'e dual of $[Q^a]$. Comparing with (\ref{gkfIIA}), we only need to perform the usual mirror symmetry exchange $\Omega \leftrightarrow e^{B+iJ}$ in order to relate both expressions.

Some other quantities are however more difficult to compare, as they arise from quite different mechanisms in type IIA and in type IIB models. A clear example is given by Yukawa couplings, which in type IIA arise from worldsheet instantons (recall fig. \ref{ws}.b)) and in type IIB can be computed from the overlap of three wavefunctions, just like for heterotic compactifications in the large volume limit. The computation of such Yukawa couplings in type IIB toroidal backgrounds was performed in \cite{cim04} (see also \cite{rs07}), where full agreement with the type IIA computation was shown. In general, agreement between type IIA and type IIB chiral field superpotentials arising from D-branes is an intense area of mathematical research (see e.g. \cite{agreement}), and lies at the very heart of Kontsevich's homological mirror symmetry conjecture \cite{Kontsevich94}.\footnote{Quite amazingly, this conjecture was formulated before D-branes were discovered.} 

This field theory approach also allows to extract important information regarding the K\"ahler metrics for charged matter. In toroidal models one basically finds good agreement with the result (\ref{KahlerIIA}). In general Calabi-Yau compactifications the wavefunction overlap is much more difficult to perform explicitly, but nevertheless one can extract non-trivial information about the K\"ahler metrics using some general scaling arguments and some mild assumptions on the K\"ahler potential \cite{ccq07}.

Finally, toroidal type IIB compactifications with magnetized D-branes also allow for exact CFT computations of elements of the effective action. If one considers general gauge bundles, and in particular those which are not of the diagonal form $F = \sum_i p_i dz^i \wedge d\bar{z}^i$ but that also admit oblique components like $dz^i \wedge d\bar{z}^j$, $i \neq j$, one is exploring D-brane configurations which go beyond the type IIB mirror of the D6-branes of the form (\ref{3cycleT6}). Such CFT computations for oblique gauge bundles have been performed in \cite{bblmr05} for the K\"ahler metrics, and in \cite{bt05} for the one-loop corrections to the gauge kinetic functions.

The construction of type IIB orientifold vacua precedes in many years to that of type IIA vacua, specially in the form of type I orbifold compactifications. Indeed, these type I vacua were constructed in \cite{typeIa} even before D-branes were discovered. The advent of D-branes and duality allowed to reinterpret those type I vacua and to construct much more (see, e.g., \cite{typeIb}), although only unmagnetized D-branes were considered. Later on, type I models based on magnetized D9-branes were constructed \cite{bgkl00,aads00}, basically following the initial proposal of \cite{Bachas95}. Finally, after ${\cal N}=1$ chiral models based on intersecting D6-branes were built, $D=4$, ${\cal N}=1$ type I models with non-trivial bundles have also been constructed \cite{lp03,dt05}.

\begin{figure}[htb]
\includegraphics[width=13cm, height=5cm]{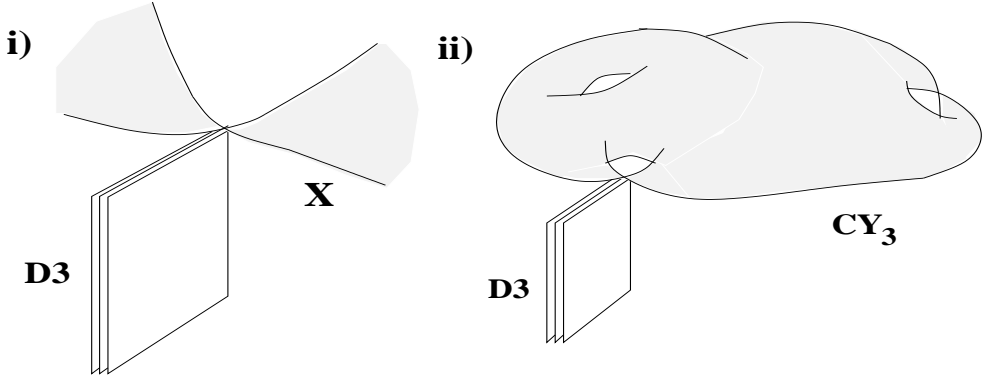}
\caption{Two-step procedure for building $D=4$ models based on D3-branes at singularities. Figure taken from \cite{aiqu00}.}
\label{bottomsup}
\end{figure}

On the other hand, type IIB vacua with O3/O7-planes have quite recently been the center of attention, specially because they can be easily embedded in type IIB flux compactifications. A particularly attractive class of models are those based on D3-branes at singularities, which were exhaustively analyzed in \cite{aiqu00} (see also \cite{bjl01}) and have been recently reconsidered in \cite{vw05}. As emphasized in \cite{aiqu00}, such D-brane models allow for a bottom-up approach when constructing them, since most of the physics regarding the gauge sector of the theory is only sensitive to local data near the singularity. One may then simply consider a local singular geometry or some blown-up version of it in order to wrap D-branes that reproduce an MSSM-like spectrum, and then complete the model by embedding such local construction into a compact Calabi-Yau (see figure \ref{bottomsup}). This construction has the clear advantage that it realizes the brane-world idea in the simplest possible way, and so it can be embedded into ${\cal N} = 0$ models with large extra dimensions and low string scale. It has also been realized that, upon decoupling of gravity, such local models allow for a realization of meta-stable vacua \cite{fu06}, embedding the field theory ideas of \cite{iss06} into a string theory setup, as well as for models of gauge mediated SUSY breaking \cite{dfks05,gsu06}. Finally, one may also consider type IIB chiral vacua based on magnetized D7-branes although, up to now, the only examples in the literature are based on toroidal backgrounds, and are T-dual to some type IIA D6-brane model, like that of table \ref{Ymodel}.

\section{Back to Type IIA \label{coisotropic}}

Given our previous discussion, one may think of type IIA and type IIB model building to be governed by the same kind of rules, as one would expect from mirror symmetry. There are however some conceptual differences between the two frameworks. For instance, in type IIB vacua we considered all kinds of space-filling D$p$-branes (i.e., those with $p \geq 3$) in order to build ${\cal N} = 1$ chiral models, while in type IIA we just considered one of them: D6-branes. Although this looked as the natural choice during our type IIA discussion, it is worth reanalyzing why the other type IIA space-filling D$p$-branes, D4 and D8-branes, were excluded from the picture. In fact, following \cite{fim06}, we will now show that D6-branes do not exhaust the set of space-filling BPS D-branes in Calabi-Yau compactifications, and that by just considering them we could be missing some big fraction of type IIA chiral vacua. 

Let us then revisit type IIA model building, and in particular the set of BPS D-branes in Calabi-Yau compactifications.  If we restrict to ${\bf CY}_3$'s with proper $SU(3)$ holonomy,\footnote{That is, all Calabi-Yau manifolds except $T^6$ and $T^2 \times {\bf K3}$.} we would never think of obtaining a BPS D-brane out of a space-filling D4-brane.  The reason is that for such manifolds $b_1 = 0$, and then our D4-brane (which is wrapping a 1-cycle of ${\bf X}_6$) would couple to a RR-potential $C_5$ without zero modes. This implies that the Chern-Simons action for a space-filling D4-brane vanishes, and so D4-branes wrapped on 1-cycles cannot carry any central charge. In practice, this is summarized by stating that D4-branes wrap 1-cycles which are homologically trivial.\footnote{Such terminology is somewhat misleading, because a D4-brane could still be wrapping a torsional 1-cycle of ${\bf X}_6$.} This obviously does not happen for a D6-brane wrapping a 3-cycle $\Pi_3 \subset {\bf X}_6$, and that is the starting point for the type IIA picture developed in section \ref{typeIIAmb}.

A similar argument to the one used for D4-branes seems to hold for D8-branes, which also wrap 5-cycles $\Pi_5 \subset {\bf X}_6$ trivial in homology. However, this is not quite true because, unlike the D4-brane, a D8-brane is allowed to carry a non-trivial gauge bundle ${\cal F}$ without breaking Poincar\'e invariance. This gauge bundle modifies the Chern-Simons action of the D8-brane and, in particular, induces a D6-brane charge on its worldvolume \cite{Douglas95}. If this D6-brane charge corresponds to a non-trivial element of $H_3({\bf X}_6, \mathbb{R})$, then we can have a non-trivial Chern-Simons action via the coupling
\begin{equation}
\int_{M_4 \times \Pi_5} {\cal F} \wedge C_7
\label{CSD8}
\end{equation}
and thus our D8-brane will have a non-trivial central charge which, just like for D6-branes, is given by an element of $H_3({\bf X}_6, \mathbb{R})$. Hence, we should be able to construct a BPS object out of a magnetized D8-brane.

From our experience with type IIA vacua it may seem quite striking that, besides D6-branes and O6-planes wrapping special Lagrangians, there could also exist stable D8-branes which are mutually BPS with the former. However, this possibility also arises in the quite different context of topological string theory. Indeed, as described in \cite{Witten92}, the prototypical example of D-brane in the topological A-model is given by a D-brane wrapping a special Lagrangian (or rather just Lagrangian) $n$-cycle of a Calabi-Yau $n$-fold. Neither the analysis in \cite{Witten92} nor the one in \cite{ooy96} exclude extra possibilities and, in fact, in \cite{ko01} it was shown that D-branes can  also wrap coisotropic $(n+2k)$-cycles, $k = 1, \dots, [n/2]$, if the appropriate worldvolume bundle $F$ is introduced. As emphasized by the authors, this fact proves essential in order to understand the full spectrum of D-branes in our theory and, on more theoretical grounds, to correctly formulate Kontsevich's conjecture.

As discussed in \cite{fim06}, our previous BPS D8-brane should be a particular case of coisotropic D-brane. This can be further argued by looking at the BPS conditions for coisotropic D-branes, which in the case of a D8-brane in a ${\bf CY}_3$ read \cite{ko01,kl03}
\begin{eqnarray}
\label{FIIAD8}
({\cal F} + iJ)^2 &  = & 0 \quad \quad {\rm F-flatness}\\
\label{DIIAD8}
{\rm Im}\, e^{-i\theta} {\cal F} \wedge \Omega & = & 0 \quad \quad {\rm D-flatness}
\end{eqnarray}
and which were used in \cite{fim06} to construct explicit examples of coisotropic D8-branes. Even in the coisotropic literature, the fact that $b_5 = 0$ in most Calabi-Yau three-folds led many to believe that stable BPS D8-branes of this kind do not exist. The examples of \cite{fim06}, however, prove the contrary.

The constructions in \cite{fim06} are based on toroidal orientifolds and, in particular, on the $\mathbb{Z}_2 \times \mathbb{Z}_2$ orientifold background used to build the MSSM-like model of table \ref{Ymodel}. One first constructs a coisotropic D8-brane on $T^2 \times T^2 \times T^2$ of the form
\begin{eqnarray}
\label{genD81}
\Pi_5 & = & (n^i,m^i)_i \times (T^2)_j \times (T^2)_k, \\
\label{genD82}
F/2\pi & = & n^{xx}\, dx^j \wedge dx^k + n^{xy}\, dx^j \wedge dy^k + n^{yx}\, dy^j \wedge dx^k + n^{yy}\, dy^j \wedge dy^k 
\end{eqnarray}
where all the $n$'s and $m$'s are integer numbers, and $\{i j k\}$ is a cyclic permutation of $\{1 2 3\}$. At this point the D-brane has non-trivial D8, D6 and D4-brane charges. Then one performs the orbifold projection, following the prescription of \cite{dm96}, and both the D8 and D4-brane charges are projected out, leaving just the D6-brane charge. In the particular case of the $\mathbb{Z}_2 \times \mathbb{Z}_2$ orbifold discussed here, one can see that no twisted charges arise when orbifolding and that the coisotropic D8-brane above is still specified by the same integer numbers, now written as
\begin{equation}
D8_a \, :\, (n^i_a, m^i_a)_i \times (n^{xx}_a, n^{xy}_a, n^{yx}_a, n^{yy}_a)_{jk}
\label{genD8}
\end{equation}
and which specify the D6-brane charge of a coisotropic D8-brane. Quite remarkably, this charge does not correspond to a 3-cycle of the form (\ref{3cycleT6}), but instead
\begin{equation}
\Pi_3^{D8}\, =\, ({\rm 1-cycle})_i \times \left\{({\rm 1-cycle})_j \times ({\rm 1-cycle})_k \oplus ({\rm 1-cycle})_j' \times ({\rm 1-cycle})_k' \right\}
\label{3cycleD8}
\end{equation}
which can in principle be carried by a D6-brane, but not by those that can be given an exact CFT description. On the other hand, the magnetized D8-brane above admits such a description, so considering these new objects allows us to enlarge our model building possibilities while staying in the `geometrical + CFT' regime of table \ref{venn}.

As shown in \cite{fim06}, many quantities of the $D=4$ effective action depend uniquely on the D6-brane charge induced on the D8-brane. In particular, notice that when we combine D6D8 and D8D8 systems, chirality arises as a combination of the intersection and magnetization mechanism, just like for type IIB D-branes. However, we do not need to compute $Ext$ groups between bundles to find the chiral spectrum, but just the usual intersection number (\ref{intersection}), where now the homology class $[\Pi_3^{D8_a}]$ coming from (\ref{3cycleD8}) must be introduced when appropriate. Hence, we can extend the D6-brane spectrum of table \ref{specIIAori} to include models with both D6 and D8-branes. The same applies to RR tadpole cancellation conditions (\ref{RRIIA}), and to the computation of the gauge kinetic function (\ref{gkfIIA}) and the FI-terms. Finally, in the case of toroidal orbifolds, other effective action elements like the K\"ahler metrics (\ref{KahlerIIA}) or one-loop corrections are also expected to be similar to those for D6-branes, since in the type I T-dual picture coisotropic D8-branes are mapped to D9-branes with oblique fluxes, and so the analyses of \cite{bblmr05,bt05} apply.\footnote{Coisotropic D8-branes can also be mapped to tilted D5 branes and magnetized D7-branes when T-dualized to type IIB with O5/O9 planes, which are objects usually not considered in the type I model building literature.}

Where coisotropic D8-branes can give different physics from D6-branes at angles is in the superpotentials that they generate. First, one can see that the F-term condition (\ref{FIIA}) is trivial for a D6-brane of the form (\ref{3cycleT6}), while the condition (\ref{FIIAD8}) is always nontrivial for a BPS D8-brane (\ref{genD8}). The superpotential corresponding to such an F-term is
\begin{equation}
W\, =\, X_i \, (T_j T_k + n)
\label{supD8}
\end{equation}
where $X_i$ is a D8-brane modulus, $T_j$ is the K\"ahler modulus of $(T^2)_j$, and $n$ comes from integrating $F^2$ over $(T^2)_j \times (T^2)_k$. While naively this superpotential leads to K\"ahler moduli stabilization, one could only analyze such issues once that all the terms in the open + closed string superpotential are taken into account. For instance, one could expect that a term of the form $W_{\rm open} = X_iX_jX_k$ also appears in the effective action, as it usually happens for D6-branes in this kind of backgrounds \cite{Douglas98}. In that case taking the F-term for the field $X_i$ would be $T_j T_k + X_j X_k + n$, and so taking the K\"ahler moduli away from the BPS point could be compensated by giving a vev to some other D8-brane moduli. In general, there is no clear, global picture of whether superpotentials generated by D-branes can lead to closed string moduli stabilization.

Second, coisotropic D8-branes also differ from D6-branes in the way that Yukawa couplings arise. Because D8-branes are not totally transverse to each other, and also do not overlap totally, Yukawa couplings may not arise from world-sheet instantons or by wavefunction overlapping, but actually by a mixture of both mechanisms. This may give rise to a richer set of patterns and textures in the fermion mass matrices than when only D6-branes were considered, although the explicit expressions for these Yukawa couplings and the new phenomenological possibilities they may add remain unexplored.

On more basic grounds, coisotropic D8-branes can also add more flexibility to the construction of explicit models. Indeed, the $\mathbb{Z}_2 \times \mathbb{Z}_2$ model of table \ref{Ymodel} is quite unique in its class and, although quite an extensive search of semi-realistic models of D6-branes has been performed for this orientifold background \cite{gbhlw05,dt06}, no simpler MSSM-like model has been found. However, by simply adding BPS D8-branes into the game one can achieve further examples of MSSM-like models, like the one presented in table \ref{D8model}, based on a combination of D6 and D8-branes.

\begin{table}[htb]
\caption{${\cal N} = 1$ type IIA vacuum based on both D6-branes and coisotropic D8-branes on $T^2 \times T^2 \times T^2/\mathbb{Z}_2 \times \mathbb{Z}_2$, also realizing the pre-model of table \ref{guay}.}
\label{D8model}
\begin{tabular}{@{}|lllc|@{}}
\hline
$D8_a$ & $N_a = 3 + 1$ & & $(1,0)_1 \times (1,3,-3,-10)_{23}$  \\
$D6_b$ & $N_b = 1$ & & $(0,1)_1 \times (1,0)_2 \times (0,-1)_3$ \\
$D6_c$ & $N_c = 1$ & & $(0,1)_1 \times (0,-1)_2 \times (1,0)_3$  \\
\hline
$D6_M$ & $N_M = 2$ & & $(-2,1)_1 \times (-3,1)_2 \times (-3,1)_3$ \\
$D8_Z$ & $N_Z = 1$ & & $(0,1)_1 \times (0,-1,-1,0)_{23}$ \\
$D6_f$ & $N_f = 4$ & & $(1,0)_1 \times (1,0)_2 \times (1,0)_3$ \\
\hline
\end{tabular}
\end{table}

Just like the model in table \ref{Ymodel}, the upper box realizes the pre-model of table \ref{guay}. Notice that the $SU(3) \times U(1)$ D6-brane has been replaced by a D8-brane, which nevertheless realizes the same intersection numbers with the other D-branes in the MSSM module. In the lower set of D-branes there are also both kinds of D-branes, yielding a total gauge group of the form
\begin{equation}
SU(4) \times SU(2)_L \times SU(2)_R \times U(2)_M \times U(1)_Z \times USp(8)
\label{gaugeD8model}
\end{equation}
where the only extra gauge group factor seen by the MSSM-like module is $U(2)_M$, by means of the extra chiral matter $6(1,2,1;2_M) + 6(1,1,2;2_M)$. One can actually get rid of this extra chiral matter, by performing a field theory Higgsing of the form $U(2)_M \rightarrow SO(2)_M$. That allows the $U(2)$ doublets to develop a mass term, and to dissappear from the massless spectrum. We are then left with just the MSSM (or rather left-right symmetric) gauge group and chiral spectrum, plus some additional gauge sector free of any chiral matter. Of course, this is not the only possibility and many more semi-realistic models may be built. For some other examples of MSSM-like models in this same orientifold background we refer the reader to \cite{fim06}.

\section{D-branes and fluxes\label{fluxes}}

Perhaps the most intense area of research in string model building during the past few years has been that of flux compactifications. From the phenomenological viewpoint, the main motivation to introduce background fluxes is that they address two main problems of Calabi-Yau vacua, moduli stabilization and supersymmetry breaking, in a controlled manner. On the other hand, they also enlarge the set of closed string backgrounds which yield $D=4$, ${\cal N} = 1$ vacua, so it is natural to analyze which are their model building possibilities. Most of the effort in the literature has been dedicated to type II strings and, in particular, to analyze which is the effect of fluxes on the closed string sector of the theory. Here we would like to focus on the effect of fluxes on D-branes and, in particular, on how they affect the model building rules described in the previous sections. 

\subsection{Type IIB fluxes}

A classical example of flux compactification is type IIB compactified on a warped Calabi-Yau and threaded by ISD background fluxes in the internal dimensions. That is, we consider a metric ansatz of the form
\begin{equation}
ds^2\, =\, \Delta^{-1}(y)\, \eta_{\mu\nu}\, dx^\mu dx^\nu\, +\, \Delta(y)\, ds^2_{{\bf CY}_3} 
\label{warped}
\end{equation}
where $y$ parameterize the six internal dimensions of a Calabi-Yau three-fold ${\bf Y}_6$, of metric $ds^2_{{\bf CY}_3}$. The warp factor $\Delta$ is sourced by the D3-branes, O3-planes and by the background 3-form flux
\begin{equation}
G_3 \, = \, F_3 - \tau H_3
\label{G3}
\end{equation}
where $F_3$, $H_3$ are the RR and NSNS 3-form field strengths, respectively, and $\tau$ is the type IIB axio-dilaton. This setup introduces a new ingredient when compared with the fluxless Calabi-Yau compactification, which is the closed string superpotential \cite{gvw99,tr99}
\begin{equation}
W_{GVW}\, = \, \int_{{\bf Y}_6} G_3 \wedge \Omega^{\textrm {\tiny CY}}
\label{GVW}
\end{equation}
where $\Omega^{\textrm {\tiny CY}}$ is the holomorphic 3-form of the unwarped Calabi-Yau ${\bf Y}_6$. Because of this superpotential the complex structure moduli and dilaton $\tau$ are lifted, and $G_3$ is forced to be an imaginary self-dual complex 3-form \cite{drs99,gkp01}. This condition is weaker than having an ${\cal N} =1$ background, which further restricts $G_3$ to have no component proportional to $\overline{\Omega}$. In addition, if ${\bf Y}_6$ contains a conifold singularity, by embedding the warped throat construction of \cite{ks00} one can naturally generate a hierarchy of $D=4$ scales \cite{gkp01}. Further research has shown that this kind of compactification provides quite an interesting framework to construct $D=4$ de Sitter vacua with small cosmological constant \cite{kklt03} and, finally, a powerful tool to better understand the vacuum structure of type II string theories \cite{add04}.

Regarding D-brane model building, a natural question is how the presence of a non-trivial $G_3$ modifies the spectrum of space-filling D-branes. From the early literature on flux compactifications and WZW models, we already know that the spectrum of D-branes before and after introducing the background flux is not the same. Indeed, by the results of \cite{as00,mms01} we know that, in the presence of $H_3$, the D3-brane charge is no longer a $\mathbb{Z}$-valued quantity, but it takes values in $\mathbb{Z}_N$ (where $N$ is the g.c.d. of $H_3$ quanta). In addition, D9-branes should be removed from the spectrum, since the Bianchi identity for the D9-brane gauge field strength ($d{\cal F} = H_3$) does not have a solution, effect usually known as Freed-Witten anomaly \cite{fw99}.

Let us now see how ${\cal N} = 1$ $G_3$ fluxes affect the D7-brane spectrum, following the discussion in \cite{gmm05}. First, one can see that the type IIB supersymmetry conditions (\ref{FIIB}) and (\ref{DIIB}) are unchanged by the presence of an ${\cal N}=1$ $G_3$ flux. In particular, if one is interested in a D7-brane wrapping a 4-cycle ${\cal S}_4 \subset {\bf Y}_6$, one should set $n=2$ and $\theta = 0$ in (\ref{DIIB}). The two conditions then imply that the gauge bundle ${\cal F}$, turned on along the internal D7-brane directions, is anti-self-dual in ${\cal S}_4$. Second, assuming a fixed closed string background $({\bf Y}_6, G_3)$ and a D7-brane $({\cal S}_4, {\cal F})$ which is initially BPS, one can see that no D-terms are generated by deforming the 4-cycle ${\cal S}_4$, basically because an ${\cal N}=1$ flux background satisfies $G_3 \wedge J = 0$.

This leaves the supersymmetry conditions (\ref{FIIB}). For simplicity, let us first assume that the D7-brane is `unmagnetized'. More precisely, we assume that the $U(1)$ gauge bundle $F = dA = 0$ and so the gauge invariant field strength ${\cal F} = 2\pi \alpha' F + B$ reduces to the pull-back of the B-field on ${\cal S}_4$. Then, for $G_3 = 0$, the moduli space of a BPS D7-brane is given by the deformations of ${\cal S}_4$ that keep it holomorphic and the deformations of the gauge potential $A$ that do not change ${\cal F}$. This translates into $h^{0,2}$ complex geometric moduli $\{\zeta^a\}$ and $h^{0,1}$ complex Wilson lines $\{\xi^b\}$ \cite{jl04}. When $G_3$ is present things change, because now ${\cal F}$ depends on the D7-brane position via the Bianchi identity $d{\cal F} = H_3$. Some geometric deformations will preserve the BPS condition ${\cal F}^{(2,0)} = 0$ but some will not, the latter being lifted D7-brane moduli. It turns out that the number of constraints imposed by ${\cal F}^{(2,0)} = 0$ is generically the same as the number of geometric moduli. Indeed, we find that the geometric moduli space of a D7-brane is given by
\begin{equation}
\begin{array}{rcl}
a_1(\zeta^1, \dots, \zeta^{h^{0,2}}) & = & 0\\
& \vdots & \\ 
a_{h^{0,2}}(\zeta^1, \dots, \zeta^{h^{0,2}}) & = & 0
\end{array}
\quad \quad \quad
\partial_{\zeta^i} a_j\, =\, \int_{{\cal S}_4} \iota_{X_i} \Omega^{\textrm {\tiny CY}} \wedge \iota_{X_j}\bar{G}_3
\label{moduliD7}
\end{equation}
where $\iota_{X_i} \alpha_p$ stands for a $p$-form $\alpha_p$ contracted with the infinitesimal holomorphic deformation $X_i$ of ${\cal S}_4$. Because by supersymmetry $G_3$ is a $(2,1)$-form, one can also see that $\partial_{\bar{\zeta}_i} a_j = 0$ and hence the $a_i$'s are holomorphic functions of the complex geometric moduli $\{\zeta^a\}$. Thus, generically (\ref{moduliD7}) is a saturated system of holomorphic equations and all the geometric moduli of a D7-brane are lifted by the presence of $G_3$.

One can easily check that this discussion also applies when we consider more general D7-branes with $F\neq 0$. In addition, in this case one may get extra sources of moduli stabilization, which may lift some D7-brane geometric moduli even in the absence of $G_3$ background fluxes \cite{Denef,km06}. On the other hand, one can see that the Wilson line moduli $\{\xi^b\}$ are not lifted by fluxes, simply because the BPS equations (\ref{FIIB}) do not depend on them (and that is why they did not appear in (\ref{moduliD7})). However, we do not even expect them to be there, since for a generic 4-cycle ${\cal S}_4$ on a Calabi-Yau we have that $b_1({\cal S}_4) = 0$. Hence, in generic flux compactifications all D7-brane moduli should be lifted. 

The BPS equations on the D7-brane moduli space can be given a $D=4$ interpretation in terms of an effective superpotential $W_{D7}$ for the D7-brane moduli. Indeed, one can see each of the equations in (\ref{moduliD7}) as the $D=4$ supersymmetry conditions
\begin{equation}
a_i \, =\, \partial_{\zeta^i} W_{D7} \, = \ 0
\label{superD7}
\end{equation}
where $W_{D7}$ has been derived in \cite{jl05} in the absence of background fluxes, and later on in \cite{Martucci06} for general ${\cal N} = 1$ backgrounds. One should also be able to derive this superpotential by considering an F-theory lift of the above setup, interpreting the Gukov-Vafa-Witten superpotential in F-theory $W = \int_{{\bf Y}_8} G_4 \wedge \Omega_4$ as a superpotential for the D7-brane moduli. Such a derivation of the D7-brane superpotential is in principle more powerful than the one followed above, since it does not need to treat the D7-brane as a probe of the closed string background. However, the geometry involved in F-theory compactifications on Calabi-Yau four-folds is in general quite complicated, and the connection with $W_{D7}$ has so far only been made for F-theory compactifications on ${\bf K3} \times {\bf K3}$ \cite{gktt04,lmrs05}. 

\begin{figure}[tb]
\includegraphics[width=13cm, height=6.125cm]{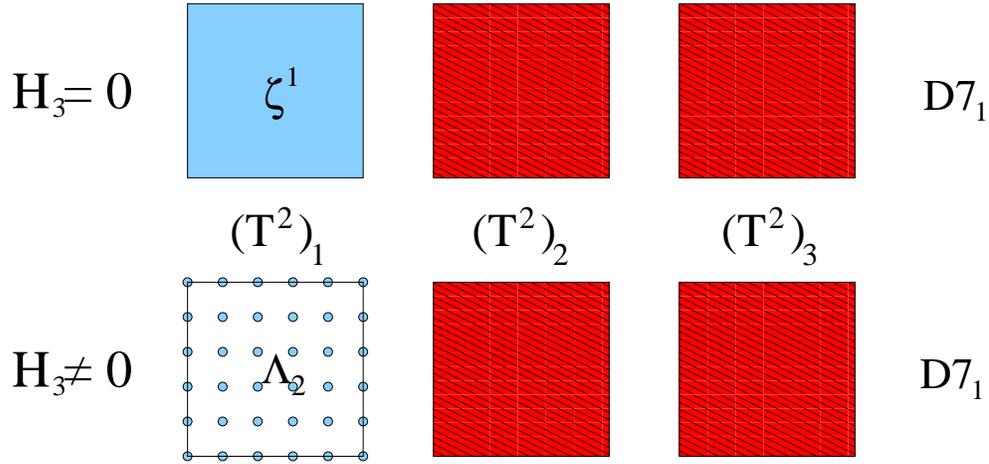}
\caption{Emergence of the D7-brane discretum.}
\label{osl}
\end{figure}

Now, an important point in this discussion is to realize that (\ref{moduliD7}) is a local description of the D7-brane moduli space, and that it does not give much information about global aspects. For instance, the fact that geometric moduli are lifted does not tell us whether the moduli space of a D7-brane is a single point or a lattice. In \cite{gmm05} it was argued that, in general, BPS D7-branes form a disconnected and discrete set of solutions to (\ref{FIIB}), and so the familiar type IIB closed string landscape should be extended to the open string sector. The simplest example of such an `open string landscape' is given by a D7-brane on $T^4 \subset T^6$, as illustrated in figure \ref{osl}. In the absence of background fluxes, the D7-brane has one complex geometric modulus $\zeta^1$, which is its position in the transverse directions to its worldvolume. When a 3-form flux $H_3$ is introduced, the pull-back of the B-field on the D7-brane will vary continuously with $\zeta^1$, and will in general spoil the BPS condition (\ref{FIIB}). One can then compensate this non-BPS B-field by means of a non-trivial gauge bundle $F$, since only the sum ${\cal F} = 2\pi\alpha' F + B$ appears in the BPS equations. However, and because $F$ is quantized, this can only be done for a discrete set of values of $\zeta^1$: those lying in the lattice $\Lambda_2$.

Another very interesting effect of fluxes on D-branes is that of generating soft SUSY breaking terms in their worldvolume action \cite{Grana02,ciu03,ggjl03}. A general analysis of type IIB vacua based on warped Calabi-Yaus shows that ISD fluxes will not lift any moduli or generate any soft term for space-filling D3-branes, while they will be indeed generated for anti-D3-branes. On the other hand, we have already seen that D7-brane moduli are quite sensitive to fluxes, so one would also expect them to be more sensitive to SUSY breaking via ISD fluxes. Indeed, it was shown in \cite{ciu04} that for D7-branes in flat space or toroidal orbifolds, ${\cal N}=0$ ISD fluxes generate a rich pattern of soft terms for chiral matter, which incidentally present interesting phenomenological features \cite{Ibanez04}. These results are based on a DBI analysis of the D7-brane worldvolume theory, and in this sense they represent a stringy way to generate soft terms in the gauge sector of the theory. For D7-branes with Abelian gauge groups one can generalize them to general warped Calabi-Yau geometries \cite{gmm05,camara06} and, in the case of small warping, they can be obtained from a $D=4$ effective action analysis \cite{lrs04a}.

Regarding the explicit chiral type IIB flux models, there are some examples in the literature although, for reasons to be discussed below, the number of ${\cal N}=1$ semi-realistic constructions is much lower than in the fluxless case. A quite natural approach to combine a semi-realistic gauge sector with all the new features introduced by fluxes is to consider models of D3-branes at singularities \cite{cgqu03,csu05}. These models can be constructed locally from an orbifold singularity, and then embedded into a warped throat via engineering an appropriate local Calabi-Yau geometry. While no soft terms would be generated for D3-branes at singularities, one can use the fact that the D-branes are at the tip of a warped throat and also consider models based on anti-D3-branes, introducing an important ingredient of the scenario proposed in \cite{kklt03}. In general, the hard part of these constructions is to find a compact Calabi-Yau manifold into which such warped throats can be embedded. 

Another approach is to build chiral models via magnetized D7-branes, which is in fact what one obtains once that a singularity with fractional D3-branes has been blown-up. The explicit models in the literature are however not based on resolved or deformed singularities, but rather on toroidal orbifolds similar to those analyzed in the intersecting D6-brane literature \cite{blt03,cu03}. In particular, one can consider the type IIB dual of the MSSM-like construction of table \ref{Ymodel}, and then combine it with the type IIB background fluxes discussed above. The MSSM-like sector of the theory will then come from magnetized D7-branes, and so the general results regarding moduli stabilization and soft SUSY breaking terms will apply (see \cite{lrs04b,fi04} for a detailed analysis of these soft terms). On the other hand, such toroidal geometries do not present any warped throat, and so it is difficult to combine such models with large warp factors. It is thus fair to say that, while much progress has been achieved, there is still no explicit semi-realistic construction that combines all the interesting features of type IIB flux vacua.

\subsection{Type IIA fluxes}

While still less developed than the type IIB framework described above, the research on type IIA flux vacua has already given some interesting results. One may then also wonder what is the interplay between D-branes and fluxes in this case. Naively one would say that, because of mirror symmetry, we should recover the same kind of effects that we observed for type IIB compactifications. One should however be careful at this point, because our analysis so far has been largely based on the geometrical intuition gained in the supergravity regime, and it is known that the mirror of a type IIB flux compactification may well be of non-geometric nature. We will here restrict to type IIA flux vacua which admit a global geometrical description, leaving some brief comments about non-geometric backgrounds for the end of this section.

Given this setup, and if we restrict to $D=4$ Poincar\'e invariant type IIA flux vacua, one should be able to reproduce the physics observed in the type IIB case. Now, the easiest way to see how type IIA reproduces the interplay between D-branes and fluxes observed above is to analyze mirror pairs of compactifications, both with geometric description. By the results of \cite{glmw02,gmpt04}, we know that the type IIB flux backgrounds above should then correspond to type IIA compactified on half-flat, symplectic manifolds ${\bf X}_6$ with $H_3 = 0$. In addition, the topological information carried by $H_3$ in the type IIB picture should be mapped to the torsion factors in the cohomology groups $H^n({\bf X}_6, \mathbb{Z})$ \cite{Tomasiello05}. This torsional cohomology ${\rm Tor}\, H^n({\bf X}_6, \mathbb{Z})$ is invisible to de Rham cohomology, and hence to usual Kaluza-Klein reduction. On the other hand, notice that all the effects produced on type IIB D-branes are due to the presence of the NSNS flux $H_3$, rather than to its RR companion $F_3$. Hence, we should expect to understand the same kind of effects in type IIA by looking at the torsion factors ${\rm Tor}\, H^n({\bf X}_6, \mathbb{Z})$.

In fact, a great deal of intuition can be gained by looking at the explicit mirror pairs constructed in \cite{kstt02}, which are based on toroidal type IIB flux compactifications and type IIA compactified on nilmanifolds or twisted tori $\tilde{T}^6$. For instance, the type IIB background
\begin{eqnarray}
\label{exmetricIIB}
ds^2 & = & (dx^1)^2 + (dx^2)^2 + (dx^3)^2 + (dx^4)^2 + (dx^5)^2 + (dx^6)^2 \\
\label{exfluxIIB}
H_3 & = & N\, dx^1 \wedge dx^5 \wedge dx^6 - N\, dx^4 \wedge dx^2 \wedge dx^6
\end{eqnarray}
is mapped to the type IIA background
\begin{equation}
\label{exmetricIIA}
ds^2  =  (dx^1 - Nx^6dx^5)^2 + (dx^2 + Nx^4dx^6)^2 + (dx^3)^2 + (dx^4)^2 + (dx^5)^2 + (dx^6)^2
\end{equation}
with $H_3 = 0$. The extra terms that appear for $N \neq 0$ in (\ref{exmetricIIA}) can be understood as a geometric `twisting' of the toroidal metric, and are also dubbed geometric fluxes. One can see that the effect of this twisting is to reduce the de Rham cohomology with respect to $T^6$ and to produce torsion factors ${\rm Tor}\, H^n(\tilde{T}^6, \mathbb{Z})$, as illustrated in table \ref{torsion} for the example at hand.
\begin{table}[htb]
\caption{Cohomology groups of the torus $T^6$ and twisted torus $\tilde{T}^6$ examples of (\ref{exmetricIIB}) and (\ref{exmetricIIA}), respectively.}
\label{torsion}
\begin{tabular}{@{}l|cccc@{}}
\hline
& $b^n(T^6)$ & ${\rm Tor}\, H^n(T^6, \mathbb{Z})$ & $b^n(\tilde{T}^6)$ & ${\rm Tor}\, H^n(\tilde{T}^6, \mathbb{Z})$ \\
\hline
n=1 & 6 & - & 4 & - \\
n=2 & 15 & - & 9 & $\mathbb{Z}_N^2$ \\
n=3 & 20 & - & 12 & $\mathbb{Z}_N^4$ \\
n=4 & 15 & - & 9 & $\mathbb{Z}_N^4$ \\
n=5 & 6 & - & 4 & $\mathbb{Z}_N^2$ \\
\hline
\end{tabular}
\end{table}

Looking at this simple example one can now understand the effects of fluxes on D-branes from the type IIA perspective \cite{Marchesano06}. Indeed, upon mirror symmetry, a D3-brane in the type IIB flux background (\ref{exmetricIIB}) is mapped to a D6-brane wrapped on a torsional 3-cycle, that is, to an element of ${\rm Tor}\, H^3(T^6, \mathbb{Z})$, and that is why its RR charge is $\mathbb{Z}_N$-valued. In a Calabi-Yau compactification such a D6-brane would never be BPS, but in the larger class of half-flat, symplectic manifolds this is possible because in general $d\Omega \neq 0$. On the other hand, the mirror of a D9-brane is a D6-brane wrapping a non-closed 3-chain, and so the Freed-Witten anomaly takes a purely geometrical nature  in the type IIA setup.\footnote{This can also be derived by imposing gauge invariance on the $D=4$ superpotential \cite{cfi05} or by means of localized BI's \cite{vz06}.}

With respect to D7-branes, they will be mapped to D6-branes which wrap usual 3-cycles, so one would like to understand how moduli stabilization by fluxes works here. Just like for type IIB D-branes, a BPS D6-brane in a geometric type IIA flux background should still satisfy the supersymmetric equations (\ref{FIIA}) and (\ref{DIIA}) \cite{ms05}. That is, it should wrap a `special Lagrangian' submanifold $\Pi_3$, even if $\Omega$ is no longer a calibration 3-form, and much of the intuition of calibrated geometry is lost. The fact that ${\bf X}_6$ is a symplectic, half-flat manifold implies that $dJ = d{\rm Im}\, \Omega = 0$, which in turn means that the results of \cite{McLean98} still apply to this case. Hence, we again have that a D6-brane in a geometric type IIA flux compactification to Minkowski has $b_1(\Pi_3)$ complex moduli. Such moduli are subject to a superpotential generated by world-sheet instantons but, in general, we do not expect much difference for D6-brane moduli stabilization with respect to the Calabi-Yau compactifications of section \ref{typeIIAmb}.

This may look quite striking, because on the one hand type IIB fluxes dramatically change the moduli space of a D7-brane, and on the other hand type IIA fluxes seem not to affect the D6-brane picture. However, as argued in \cite{Marchesano06}, there is no contradiction between these facts and the known examples of mirror pairs of flux vacua in the literature, like those discussed above. As a result, when building type IIA flux vacua based on intersecting D6-branes, one cannot rely on fluxes to lift open string moduli and should use complementary strategies like, e.g., models based on rigid 3-cycles (i.e., those with $b_1(\Pi_3) = 0$) \cite{bcms05}, ${\cal N} = 0$ backgrounds based on Scherk-Schwarz orientifolds \cite{aci05}, etc.

Building explicit models based on type IIA flux vacua is still more difficult than for type IIB, basically because the compactification ansatz is not so simple as in (\ref{warped}) (see \cite{abv06} for progress on this issue). Hence the examples in the literature are based on toroidal orientifolds and close relatives, where the warp factor is mild and can be ignored. Such examples were initially based on type IIA AdS vacua without D6-branes \cite{dgkt05}, and later on Minkowski and AdS vacua including D6-branes were constructed \cite{cfi05}. The main advantage which these models offer is a greater flexibility for moduli stabilization, since by using different kinds of fluxes (more than those used in the type IIB framework above) one can basically stabilize all closed string moduli of these compactifications. In addition, fluxes relax the RR tadpole cancellation conditions, which (naively) translates into a greater flexibility when building consistent D-brane models.

\subsection{Non-geometric backgrounds}

As stated before, the mirror of a type IIB flux background may not have a geometric description.\footnote{More precisely, if a type IIB background of the form (\ref{warped}) contains $H_3$ fluxes which are `magnetic' in the natural symplectic basis of 3-forms of a Calabi-Yau manifold, one expects the mirror type IIA background to be given by an $SU(3) \times SU(3)$-structure manifold, which generically corresponds to a non-geometric compactification \cite{glw06}.} While intuitively less clear, one can still make sense of these backgrounds as string theory compactifications \cite{Hull04}, and also compute $D=4$ effective theory quantities like a generalized superpotential for closed string moduli \cite{stw05}. Finally, duality arguments suggest that the different kinds of `non-geometric fluxes' that one may introduce in type II theories form a much larger class than the geometric ones \cite{acfi06}. While our knowledge of these non-geometric constructions is still quite poor, and mainly based on mirrors of toroidal compactifications, the above results have led many to believe that non-geometric backgrounds correspond to the largest fraction of the `type II landscape'. A fraction which has so far been unexplored.

The main features of these non-geometric compactifications highlighted so far follow the trend of the type IIA flux backgrounds. As we add more fluxes we have a more general flux-induced superpotential, which allows to stabilize more closed string moduli. In addition, the larger the amount of fluxes is, the more the RR tadpole conditions can be relaxed. Again, this naively indicates that, when adding more and more fluxes, it will become easier to build semi-realistic models, as we can basically modify the tadpole constraints to our will. One should however be careful when making such a claim because, as we have seen during our discussion, the interplay between D-branes and fluxes is non-trivial and, in general, the inclusion of background fluxes modifies the D-brane spectrum of a compactification. How this works for non-geometric vacua is still not known, since we do not have a full understanding of the D-brane spectrum in this case (see \cite{lsw06,vz06,km06} for some results on this problem). We can however learn some lesson from the effects of geometric fluxes described above.

Indeed, one of the main effects of background fluxes is that they produce torsional RR charges in a compactification. The simplest example of this is given by (non-fractional) D3-branes in warped Calabi-Yau compactifications with ISD background fluxes, or their type IIA mirror D6-branes wrapping torsional 3-cycles. Now, from the type IIA picture it is easy to see that a D6-brane on a torsional 3-cycle cannot create chirality, in the sense that any intersection number (\ref{intersection}) involving such a D-brane will automatically vanish \cite{Marchesano06}. It could then happen that by introducing more and more kinds of background fluxes in a compactification more and more D-brane charges become torsional, and hence obtaining a chiral spectrum rich enough to embed the MSSM could become more and more difficult. While there are still no no-go theorems in this aspect of flux compactifications, it is clear that some tension may exist between fluxes and chirality, and this is something to have in mind when trying to build semi-realistic string vacua.

\section{Final comments}

In this note we have briefly reviewed recent ideas and developments in D-brane model building, first in standard Calabi-Yau orientifold compactifications and then in more novel flux vacua. In Calabi-Yau compactifications we have first focused on the model building rules to construct ${\cal N} = 1$, $D=4$ vacua with semi-realistic gauge group and chiral matter content, and later considered the computation of effective field theory quantities like gauge kinetic functions, Yukawa couplings, etc. In this quest, we have benefited from the intuitive picture that arises in intersecting D6-brane models, which perhaps provide the conceptually simplest framework to describe chiral spectra in string theory. This simplicity reflects on the large amount of semi-realistic models that have been built in this framework, and on how this type IIA picture has influenced other model building areas like those of type IIB, heterotic \cite{bhw05a} and M-theory \cite{csu01b}. Despite the intense research carried out over the past few years, we have seen that some elements of type IIA model building have been missing until very recently, like the fact that D8-branes can also be building blocks of ${\cal N}=1$ chiral vacua. We have seen some model building applications of these new objects in a simple background, but what is their role in the whole type IIA picture or whether new BPS objects remain to be discovered are still open questions.

In flux compactifications we basically rely on the same D-brane mechanisms to create chirality as in the fluxless case, but somehow the model building rules change because the spectrum of D-branes is directly affected by fluxes. We have analyzed three main effects of fluxes on D-branes: the Freed-Witten anomaly that removes some D-branes, the fact that fluxes render some D-brane charges torsional, and the effect of open string moduli stabilization/D-brane discretum. We have described such effects in relatively well-understood type IIA and type IIB vacua by using basic arguments, but more involved techniques, like those based on generalized calibrations, may give further insight when analyzing more complicated compactifications \cite{cu04,km06}. 

In addition, fluxes affect the D-brane effective theories by inducing supersymmetry breaking soft terms. This is a very promising avenue in string model building, although also quite involved because in actual compactifications we need to take into account the non-trivial interplay between closed and open string sectors of the theory. For instance, one may conceive breaking supersymmetry by type IIB ISD fluxes, but then the induced D-brane soft terms will depend on the vev's of the closed string fields, so they need to be stabilized in order to have specific soft terms. The process of closed string moduli stabilization usually involves superpotentials created from D-branes \cite{kklt03}, which may affect the soft terms and are in turn affected by the background fluxes. So we end up with quite a complicated coupled system, which should nevertheless be self-consistent like in \cite{lrst06}. It is however clear that we need to understand the dynamics of all the relevant sectors of the theory before making claims about moduli stabilization and supersymmetry breaking. In particular, one may wonder if space-filling D-branes like the ones analyzed here have a relevant role in closed string moduli stabilization, once that we stop treating them as probes. Several proposals and scenarios have been suggested in this respect \cite{proposals}, although it is fair to say that no global picture is yet available. 

Finally, let us point out that we have mainly focused on space-filling D-branes in our discussion, and have ignored the role played by instantonic or Euclidean D-branes. As recently noticed in the literature these may not only affect the closed string sector of the theory via a non-perturbative superpotential \cite{superinst}, but also the open string sector. The relevant Euclidean D-branes in the latter case are those which intersect the space-filling D-branes of a given model, generating a non-perturbative superpotential for the chiral matter fields. The phenomenological consequences of these instanton effects have been recently analyzed in \cite{bcw06,iu06,fkms06} (see also \cite{ag06,bk07}), with the result that they provide new sources for moduli stabilization, generation of $\mu$-terms and neutrino Majorana masses. Just like all the other developments described in this review, instantonic D-branes represent a piece of progress in approaching string theory to particle physics. Our hope is that, when combined with previous and future model building ideas, we can gain some insight in how string theory may realize our universe. Maybe then we will understand whether a theory of strings can solve the major contemporary puzzles of fundamental Physics, and live up to its mid 80's expectations.

\begin{acknowledgement}
I would like to thank the organizers of the RTN Workshop and Midterm meeting 2006 in Napoli for their invitation to give a review talk, as well as the Instituto de Astrofisica y Fisica del Espacio (IAFE) and the Instituto Balseiro, Centro At\'omico de Bariloche (Argentina) for hospitality while this report was written. It is also a pleasure to thank L.~Ib\'a\~nez, D.~Krefl, A.~Uranga and specially M.~Haack for useful comments on the manuscript. This work is supported by the European Network ``Constituents, Fundamental Forces and Symmetries of the Universe", under the contract MRTN-CT-2004-005104.
\end{acknowledgement}

\end{document}